\pdfoutput=1

\documentclass[11pt,twoside,a4paper,cmspaper,final,collab]{cms-tdr}
\def\svnVersion{5db41e0-D}\def\svnDate{2019/06/22}\def\cmsCernNoTag{CERN-EP-2018-139}\def\cmsCernDate{\today}\def\cmsMessage{"Published in Physics Letters B as \href{http://dx.doi.org/10.1016/j.physletb.2019.04.025}{\doi{10.1016/j.physletb.2019.04.025}.}"}

\begin{document}\cmsNoteHeader{HIG-17-023}

\hyphenation{had-ron-i-za-tion}
\hyphenation{cal-or-i-me-ter}
\hyphenation{de-vices}
\RCS$HeadURL$
\RCS$Id$

\newlength\cmsFigWidth
\ifthenelse{\boolean{cms@external}}{\setlength{\cmsFigWidth}{\columnwidth}}{\setlength{\cmsFigWidth}{\textwidth}}
\ifthenelse{\boolean{cms@external}}{\providecommand{\cmsLeft}{upper\xspace}}{\providecommand{\cmsLeft}{left\xspace}}
\ifthenelse{\boolean{cms@external}}{\providecommand{\cmsRight}{lower\xspace}}{\providecommand{\cmsRight}{right\xspace}
\providecommand{\CL}{CL\xspace}}

\newcommand{\PV}{\ensuremath{\cmsSymbolFace{V}}\xspace}
\newcommand{\Vjets}{\ensuremath{\PV\text{+jets}}\xspace}
\newcommand{\Zmm}{\ensuremath{\PZ(\mu\mu)}\xspace}
\newcommand{\Zee}{\ensuremath{\PZ(\Pe\Pe)}\xspace}
\newcommand{\Zll}{\ensuremath{\PZ(\ell\ell)}\xspace}
\newcommand{\Zvv}{\ensuremath{\PZ(\nu\overline{\nu})}\xspace}
\newcommand{\Wlv}{\ensuremath{\PW(\ell\nu)}\xspace}
\newcommand{\Wmn}{\ensuremath{\PW(\mu\nu)}\xspace}
\newcommand{\Wen}{\ensuremath{\PW(\Pe\nu)}\xspace}
\newcommand{\WCR}{\PW(CRs)\xspace}
\newcommand{\ZCR}{\PZ(CRs)\xspace}
\newcommand{\MINLO} {{\textsc{minlo}}\xspace}
\newcommand{\Zmmjets}{\ensuremath{\PZ(\mu\mu)\text{+jets}}\xspace}
\newcommand{\Zeejets}{\ensuremath{\PZ(\Pe\Pe)\text{+jets}}\xspace}
\newcommand{\Zlljets}{\ensuremath{\PZ(\ell\ell)\text{+jets}}\xspace}
\newcommand{\Zjets}{\ensuremath{\PZ\text{+jets}}\xspace}
\newcommand{\Wjets}{\ensuremath{\PW\text{+jets}}\xspace}
\newcommand{\Zvvjets}{\ensuremath{\PZ(\PGn\PAGn)\text{+jets}}\xspace}
\newcommand{\Wlvjets}{\ensuremath{\PW(\ell\nu)\text{+jets}}\xspace}
\newcommand{\Wmnjets}{\ensuremath{\PW(\mu\nu)\text{+jets}}\xspace}
\newcommand{\Wevjets}{\ensuremath{\PW(\Pe\nu)\text{+jets}}\xspace}
\newcommand{\phojets}{\ensuremath{\gamma\text{+jets}}\xspace}
\newcommand{\kappaz}{\ensuremath{\kappa\smash[b]{^{\PGn\PAGn}}}\xspace}
\newcommand{\kappaV}{\ensuremath{\kappa_{\mathrm{V}}}\xspace}
\newcommand{\kappaF}{\ensuremath{\kappa_{\mathrm{F}}}\xspace}
\newcommand{\kappazi}{\ensuremath{\kappa_{i}\smash[b]{^{\PGn\PAGn}}}\xspace}
\newcommand{\brhinv}{\ensuremath{{\mathcal{B}(\PH \to \text{inv})}}\xspace}
\newcommand{\hinv}{\ensuremath{\PH \to \text{inv}}\xspace}
\newcommand{\mettrig}{\ensuremath{p_{\text{T, trig}}^{\text{miss}}}\xspace}
\newcommand{\mhttrig}{\ensuremath{H_{\text{T, trig}}^{\text{miss}}}\xspace}
\newcommand{\mjj}{\ensuremath{m_{\mathrm{jj}}}\xspace}
\newcommand{\detajj}{\ensuremath{\Delta\eta_{\mathrm{jj}}}\xspace}
\newcommand{\dphijj}{\ensuremath{\Delta\phi_{\mathrm{jj}}}\xspace}
\newcommand{\dphijmet}{\ensuremath{\Delta\phi(\ptvecmiss,\vec{p}_{\mathrm{T}}^{\kern1pt\text{jet}})}\xspace}
\newcommand{\sigmabr}{\ensuremath{(\sigma/\sigma_{\mathrm{SM}}) \, \brhinv}\xspace}
\newcommand{\HRES} {\textsc{hres}\xspace}
\newcommand{\NNPDF} {\textsc{nnpdf}\xspace}

\newlength\cmsTabSkip\setlength{\cmsTabSkip}{0.5ex}
\ifthenelse{\boolean{cms@external}}{\providecommand{\cmsTable}[1]{#1}}{\providecommand{\cmsTable}[1]{\resizebox{\textwidth}{!}{#1}}}
\providecommand{\NA}{\ensuremath{\text{---}}}

\cmsNoteHeader{HIG-17-023}
\title{Search for invisible decays of a Higgs boson produced through vector boson fusion in proton-proton collisions at $\sqrt{s} = 13\TeV$}

\date{\today}

\abstract{
A search for invisible decays of a Higgs boson is performed using proton-proton collision data collected with the CMS detector at the LHC in 2016 at a center-of-mass energy $\sqrt{s} = 13\TeV$, corresponding to an integrated luminosity of 35.9\fbinv. The search targets the production of a Higgs boson via vector boson fusion. The data are found to be in agreement with the background contributions from standard model processes. An observed (expected) upper limit of 0.33\,(0.25), at 95\% confidence level, is placed on the branching fraction of the Higgs boson decay to invisible particles, assuming standard model production rates and a Higgs boson mass of 125.09\GeV. Results from a combination of this analysis and other direct searches for invisible decays of the Higgs boson, performed using data collected at $\sqrt{s}=7$, 8, and 13\TeV, are presented. An observed (expected) upper limit of 0.19\,(0.15), at 95\% confidence level, is set on the branching fraction of invisible decays of the Higgs boson. The combined limit represents the most stringent bound on the invisible branching fraction of the Higgs boson reported to date. This result is also interpreted in the context of Higgs-portal dark matter models, in which upper bounds are placed on the spin-independent dark-matter-nucleon scattering cross section.
}

\hypersetup{
pdfauthor={CMS Collaboration},
pdftitle={Search for invisible decays of a Higgs boson produced through vector boson fusion at sqrt(s) = 13 TeV},
pdfsubject={CMS},
pdfkeywords={CMS, physics, Higgs, VBF, invisible decays}}

\maketitle

\section{Introduction}

Since the discovery of the Higgs boson at the CERN LHC~\cite{Aad:2012tfa,Chatrchyan:2012xdj,Chatrchyan:2013lba}, the ATLAS and CMS Collaborations have pursued a wide-ranging program to study its properties and interactions. Precise measurements of the couplings of the Higgs boson to standard model (SM) particles indicate that the properties of the new particle are consistent with the SM predictions~\cite{Khachatryan:2016vau}. These measurements also provide indirect constraints on additional contributions to the Higgs boson width from beyond the SM (BSM) decays. Based on the results presented in Ref.~\cite{Khachatryan:2016vau}, an indirect upper limit on the Higgs boson branching fraction to BSM particles of 0.34 is set at 95\% confidence level (\CL).

In the SM, the Higgs boson decays invisibly (\hinv) only through the $\PH \to \PZ\PZ \to 4\nu$ process, with a branching fraction, \brhinv, of about $10^{-3}$. The rate for invisible decays of the Higgs boson may be significantly enhanced in the context of several BSM scenarios~\cite{Belanger:2001am,Datta:2004jg,Dominici:2009pq,SHROCK1982250}, including those in which the Higgs  boson acts as a portal to dark matter (DM)~\cite{Djouadi:2011aa,Baek:2012se,Djouadi:2012zc,Beniwal:2015sdl}. Direct searches for \hinv decays increase the sensitivity to \brhinv beyond the indirect constraints. The ATLAS Collaboration~\cite{Aad:2008zzm} presented a combination of direct searches using $\sqrt{s}=7$ and 8\TeV data from proton-proton ($\Pp\Pp$) collisions, yielding an observed (expected) upper limit of 0.25 (0.27) on \brhinv at 95\% \CL~\cite{Aad:2015pla}. The CMS Collaboration~\cite{Chatrchyan:2008zzk} performed a similar combination based on $\sqrt{s}=7$, 8, and 13\TeV $\Pp\Pp$ collision data collected up to the end of 2015, setting an observed (expected) upper limit of 0.24 (0.23) on \brhinv at 95\% \CL~\cite{Khachatryan:2016whc}.

\begin{figure*}[htb]
\centering
\includegraphics[width=0.30\textwidth]{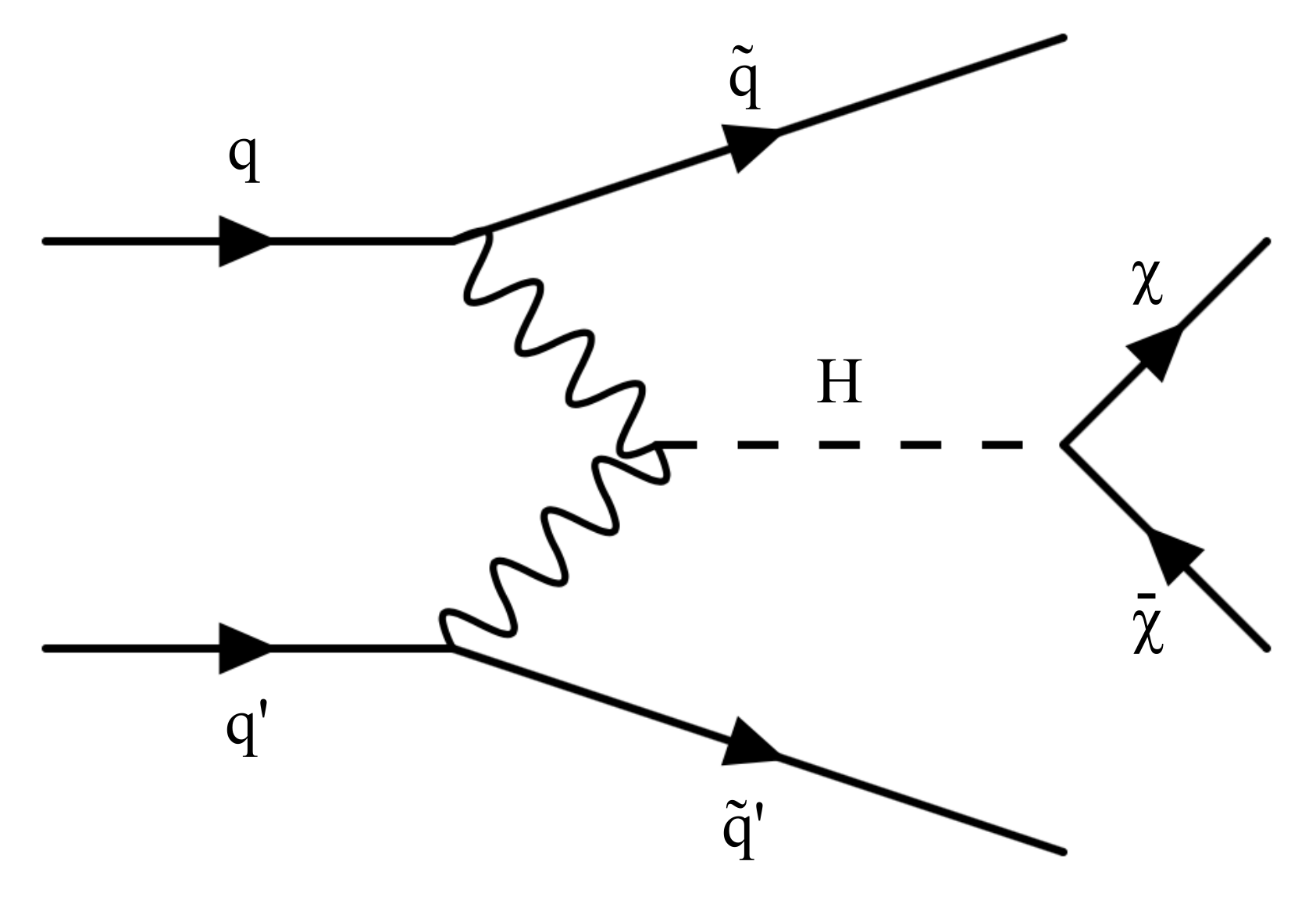}
\hspace*{\fill}
\includegraphics[width=0.30\textwidth]{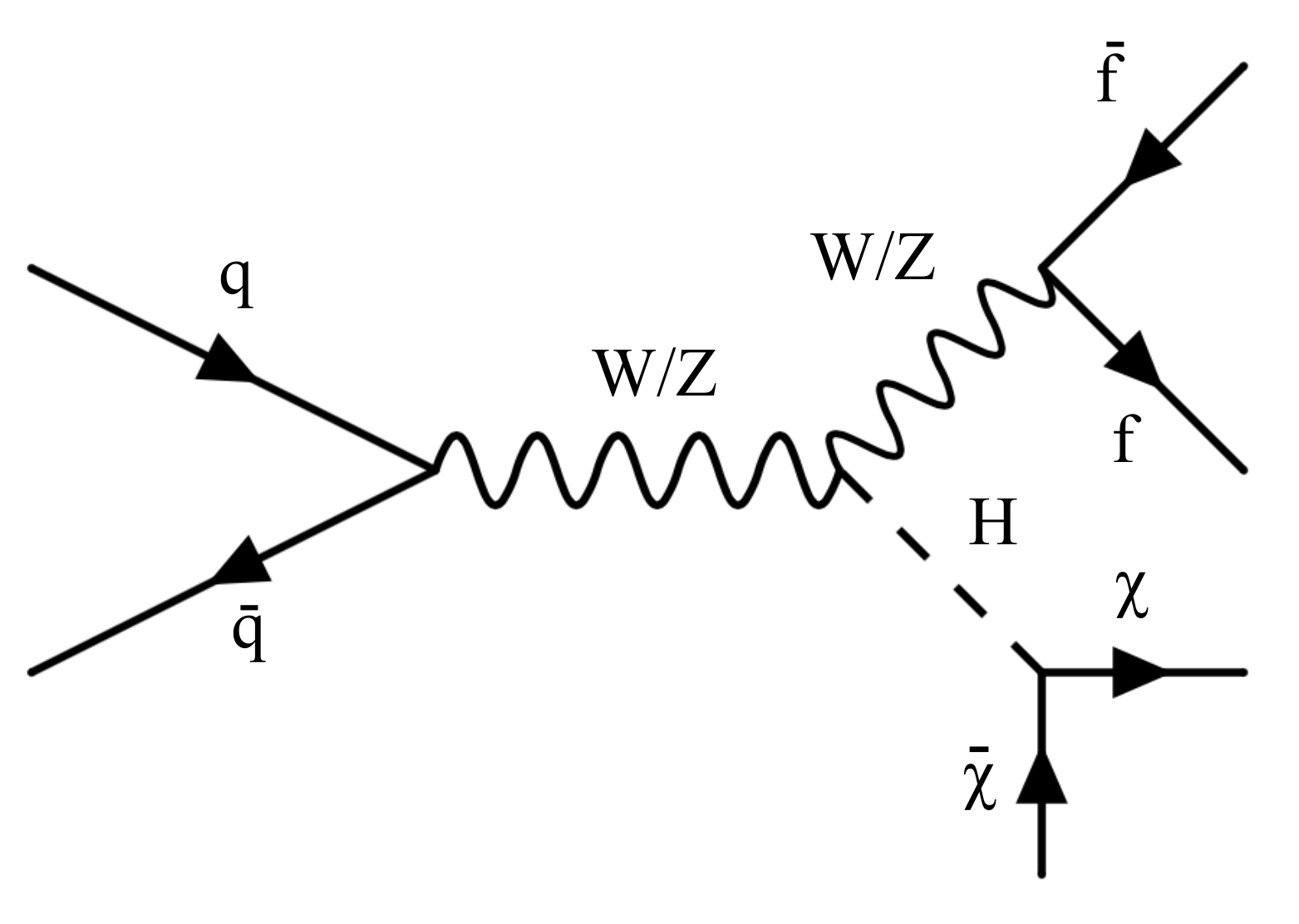}
\hspace*{\fill}
\includegraphics[width=0.30\textwidth]{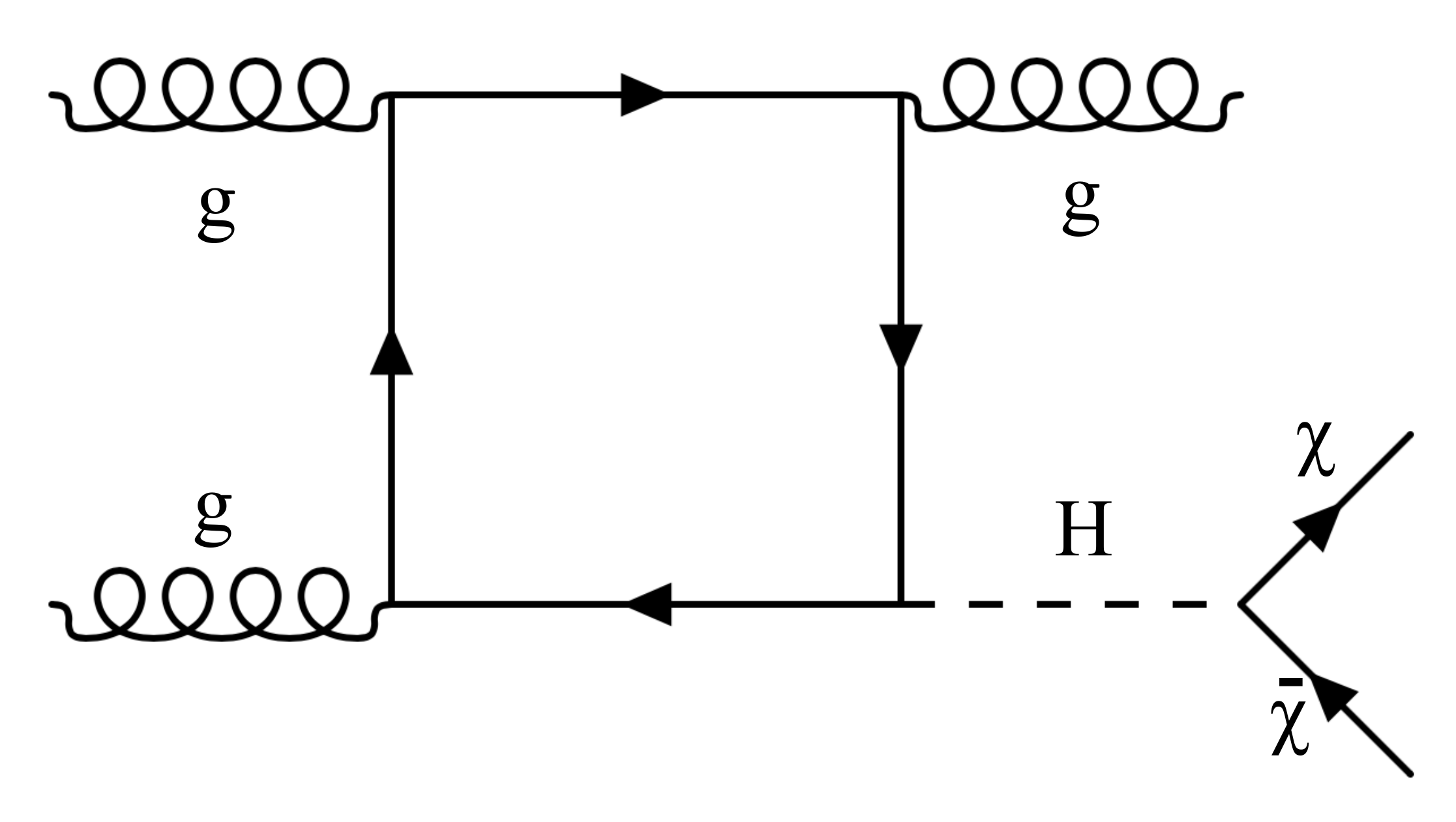}
\caption{Leading order Feynman diagrams for the main production processes targeted in the combination: VBF (left), ${\PV\PH}$ (middle), and ${\Pg\Pg\PH}$ (right).}
\label{fig:feynman_graphs}
\end{figure*}

This Letter presents a search for invisible decays of a Higgs boson, using $\Pp\Pp$ collision data at $\sqrt{s} = 13 \TeV$ collected with the CMS detector in 2016, corresponding to an integrated luminosity of 35.9\fbinv. The search targets events in which a Higgs boson is produced in association with jets from vector boson fusion (VBF), as illustrated by the Feynman diagram in Fig.~\ref{fig:feynman_graphs} (left). In these events, a Higgs boson is produced along with two jets that exhibit a large separation in pseudorapidity ($\abs{\detajj}$) and a large dijet invariant mass (\mjj). This characteristic signature allows for the suppression of SM backgrounds, making the VBF channel the most sensitive mode for invisible decays of a Higgs boson at hadron colliders~\cite{Eboli:2000ze,Khachatryan:2016whc}. The invisible particles produced by the Higgs boson decay can recoil with high transverse momentum (\pt) against the visible VBF-jet system, resulting in an event with large \pt imbalance, which can be used to select signal enriched regions. In this phase space, the main expected backgrounds originate from \Zvvjets and \Wlvjets processes. They are estimated from data using dedicated control regions (CRs), which consist of high purity samples of $\PZ$ or $\PW$ bosons decaying leptonically (${\ell = \mu,\Pe}$). While earlier searches probing this final state at the LHC were based on counting experiments, the analysis presented in this Letter more optimally exploits the distinctive kinematic features of the VBF topology by fitting the shape of the \mjj distribution. This approach is referred to as the ``shape analysis''. The shape analysis has been designed to provide a substantially improved sensitivity to invisible decays of the SM Higgs boson, resulting in the most sensitive VBF \hinv search reported to date. In addition, a simple but less sensitive counting approach, referred to as the ``cut-and-count analysis'', allows for an easier interpretation of the results of this search in the context of other phenomenological models predicting the same final-state signature. Upper limits on the product of the cross section and branching fraction for an additional Higgs boson with SM-like couplings, which does not mix with the 125\GeV Higgs boson, are also reported.

To further improve the sensitivity, results from a combination of searches for invisible decays of the Higgs boson, using data collected at $\sqrt{s} = 7,$ 8, and 13\TeV, are also presented. The searches considered in this combination target the VBF, the associated production (denoted by $\PV\PH$, where \PV denotes a $\PW$ or $\PZ$ boson), and the gluon fusion modes, whose representative Feynman diagrams are shown in Fig.~\ref{fig:feynman_graphs}. The $\PV\PH$-tag includes both a search for $\PZ\PH$ production, in which the $\PZ$ boson decays to a pair of leptons ($\Pe,\mu$) or b quarks, and one where a Lorentz-boosted $\PW$ or $\PZ$ boson decays to light-flavor quarks, whose corresponding hadronization products are reconstructed as a single large-radius jet. Additional sensitivity is achieved by including a search for ${\Pg\Pg\to\Pg\PH}$ production (hereafter referred to as $\Pg\Pg\PH$), where a high-\pt Higgs boson candidate is produced in association with jets from initial-state radiation. When these searches are combined to set an upper limit on \brhinv, SM production cross sections are assumed. The result of this combination is also interpreted in the context of Higgs-portal models of DM interactions~\cite{Djouadi:2011aa,Baek:2012se,Djouadi:2012zc,Beniwal:2015sdl}, in which the 125\GeV Higgs boson plays the role of a mediator between the SM and DM particles, thereby allowing for the possibility of producing DM candidates.

This Letter is organized as follows: after a brief description of the CMS detector in Section~\ref{sec:cmsdetector}, the event reconstruction in Section~\ref{sec:reconstruction}, and the simulated signal and background processes in Section~\ref{sec:simulation}, Section~\ref{sec:selection} is dedicated to the event selection requirements followed by a detailed description of the analysis strategy in Section~\ref{sec:background}. Section~\ref{sec:results} reports the results of the VBF search in terms of upper limits on \brhinv. Section~\ref{sec:combination_2016} reports the upper limit on \brhinv from a combination of the aforementioned searches for invisible decays of the Higgs boson based on 13\TeV data collected in 2016 while, in Section~\ref{sec:combination}, results from a more complete combination, involving also similar analyses performed on the 7 and 8\TeV data sets, are presented. The Letter is summarized in Section~\ref{sec:summary}.

\section{The CMS detector}\label{sec:cmsdetector}

The CMS detector is a multi-purpose apparatus designed to study a wide range of physics processes in both ${\Pp\Pp}$ and heavy ion collisions. The central feature of the experiment is a superconducting solenoid of 6\unit{m} internal diameter, providing a magnetic field of 3.8\unit{T} parallel to the beam direction.
Within the solenoid volume a silicon pixel and strip tracker, a lead tungstate crystal electromagnetic calorimeter (ECAL), and a brass and scintillator hadron calorimeter (HCAL) are installed, each composed of a barrel and two endcap sections. The tracker system measures the momentum of charged particles up to $\abs{\eta} = 2.5$, while the ECAL and HCAL provide coverage up to $\abs{\eta} = 3.0$. In addition, the steel and quartz-fiber Cherenkov hadron forward calorimeter extends the coverage to $\abs{\eta} = 5.0$. Muons are detected in gas-ionization chambers embedded in the steel flux-return yoke outside the solenoid, which cover up to $\abs{\eta} = 2.4$.

Events of interest are selected using a two-tiered trigger system~\cite{Khachatryan:2016bia}. The first level (L1) is composed of custom hardware processors, which use information from the calorimeters and muon detectors to select events at a rate of about 100\unit{kHz}. The second level, known as high-level trigger (HLT), is a software-based system which runs a version of the CMS full event reconstruction optimized for fast processing, reducing the event rate to about 1\unit{kHz}.

A more detailed description of the CMS detector, together with a definition of the coordinate system used and the relevant kinematic variables, can be found in Ref.~\cite{Chatrchyan:2008zzk}.

\section{Event reconstruction}\label{sec:reconstruction}

The particle-flow (PF) algorithm~\cite{Sirunyan:2017ulk} aims to reconstruct and identify each particle in an event with an optimized combination of information from the various elements of the CMS detector. The energy of photons is obtained from the ECAL measurement. The energy of electrons is determined from a combination of the momentum of the associated track at the primary interaction vertex, the energy of the corresponding ECAL cluster, and the energy sum of all bremsstrahlung photons spatially compatible with originating from the electron track. The momentum of muons is obtained from the curvature of the corresponding tracks. The energy of charged hadrons is determined from a combination of their momentum measured in the tracker and the matched ECAL and HCAL energy deposits, corrected for the response function of the calorimeters to hadronic showers. Finally, the energy of neutral hadrons is obtained from the corresponding corrected ECAL and HCAL energy.

The missing transverse momentum vector (\ptvecmiss) is computed as the negative vector \pt sum of all the PF candidates in an event, and its magnitude is denoted as \ptmiss.  Hadronic jets are reconstructed by clustering PF candidates using the anti-\kt algorithm~\cite{Cacciari:2008gp,Cacciari:2011ma}, with a distance parameter of 0.4. The reconstructed vertex with the largest value of summed physics object $\pt^2$ is taken to be the primary ${\Pp\Pp}$ interaction vertex, where physics objects correspond to the jets and the \ptmiss measured in the event. The charged PF candidates originating from any other vertex are ignored during the jet finding procedure. Jet momentum is determined as the vector sum of all particle momenta inside the jet, and is found from simulation to vary, on average, between 5 and 10\% of the true momentum over the whole \pt spectrum and detector acceptance. An offset correction is applied to jet energies to take into account the contribution from additional ${\Pp\Pp}$ interactions within the same or adjacent bunch crossings (pileup)~\cite{Cacciari:2007fd}. Jet energy corrections are derived from simulation and are confirmed with \textit{in situ} measurements of the energy balance in dijet, multijet, \phojets, and leptonically decaying \Zjets events~\cite{Khachatryan:2016kdb}. These energy corrections are also propagated to the \ptmiss calculation~\cite{Khachatryan:2014gga}.

Muon candidates, within the geometrical acceptance of the silicon tracker and muon subdetectors (${\abs{\eta} < 2.4}$), are reconstructed by combining the information from the tracker and the muon chambers~\cite{Chatrchyan:2012xi}. These candidates are required to satisfy a set of quality criteria based on the number of hits measured in the tracker and the muon system, the properties of the fitted muon track, as well as the impact parameters of the track with respect to the primary vertex of the event.

Electron candidates within $\abs{\eta} < 2.5$ are reconstructed using an algorithm that associates fitted tracks in the silicon tracker with electromagnetic energy clusters in the ECAL~\cite{Khachatryan:2015hwa}. To reduce the misidentification rate, these candidates are required to satisfy identification criteria based on the shower shape of the energy deposit, the matching of the electron track to the ECAL energy cluster, the relative amount of energy deposited in the HCAL detector, and the consistency of the electron track with the primary vertex. Because of non-optimal reconstruction performance, electron candidates in the transition region between the ECAL barrel and endcap, ${1.44 < \abs{\eta} < 1.57}$, are not considered in the analysis. Electrons identified as coming from photon conversions in the detector are discarded~\cite{Khachatryan:2015iwa}.

Identified electrons or muons are required to be isolated from hadronic activity in the event. The isolation is defined by summing the \pt of all the PF candidates within a cone of radius ${R = \sqrt{\smash[b]{(\Delta\eta)^2+(\Delta\phi)^2}} = 0.4\,(0.3)}$ around the muon (electron) track, and is corrected for the contribution of neutral hadrons from pileup interactions~\cite{Chatrchyan:2012xi,Khachatryan:2015hwa}.

Hadronically decaying $\tau$ leptons (\tauh) are identified from reconstructed jets via the hadron-plus-strip algorithm~\cite{Khachatryan:2015dfa}, that requires a subset of particles inside the jet to be consistent with the decay products of a $\tau$ lepton. In addition, the \tauh candidate must be isolated from other activity in the detector. The isolation is computed by summing the \pt of all the charged PF candidates and PF photons within a cone of radius $R=0.3$ around the jet axis. Hadronic $\tau$ leptons are selected with an average efficiency between 60 and 65\%.

\section{Simulated samples}\label{sec:simulation}

The signal and background processes are simulated using several Monte Carlo (MC) generators. Higgs boson signal events, produced through $\Pg\Pg\PH$ and VBF, are generated with \POWHEG v2.0~\cite{Nason:2004rx,Frixione:2007vw,Alioli:2010xd,Bagnaschi:2011tu,Nason:2009ai} at next-to-leading order (NLO) approximation in perturbative quantum chromodynamics (QCD). Signal events are normalized to the inclusive Higgs boson production cross sections taken from the recommendations of Ref.~\cite{deFlorian:2016spz}.
The $\Pg\Pg\PH$ production cross section is computed at next-to-next-to-NLO ($\mathrm{N}^{3}\mathrm{LO}$) precision in QCD, and at NLO in electroweak (EW) theory~\cite{Anastasiou:2016cez}.
The cross section for Higgs boson production through VBF is calculated at next-to-NLO (NNLO) in QCD, including also NLO EW corrections. The $\Pg\Pg\PH$ process is simulated using calculations in which the top quark loop is fully resolved. The \pt distribution of the Higgs boson produced via $\Pg\Pg\PH$ is reweighted to match the NNLO plus next-to-next-to-leading-logarithmic (NNLL) prediction from \HRES v2.1~\cite{deFlorian:2012mx,Grazzini:2013mca}. When upper limits are set on \brhinv for the SM Higgs boson, both $\Pg\Pg\PH$ and VBF signal events are generated assuming a Higgs boson mass of 125.09\GeV, which is consistent with the combined ATLAS and CMS measurement~\cite{Aad:2015zhl} based on 7+8\TeV data, as well as the recent CMS measurement at 13\TeV in the ${\PH \to \PZ\PZ \to 4\ell}$ channel~\cite{Sirunyan:2017exp}.

The $\PZ/\gamma^{*}(\ell^{+}\ell^{-})$+jets,~\Zvvjets, and \Wlvjets backgrounds are simulated at leading order (LO) using \MGvATNLO v2.2.2~\cite{Alwall:2014hca}, where up to four partons in the final state are included in the matrix element calculation. The background processes involving the production of a vector boson (\PV) in association with two jets exclusively through EW interactions, \ie of order $\alpha^{4}$, are simulated at LO via \MGvATNLO. In addition, the QCD multijet background is also simulated at LO using \MGvATNLO. The \ttbar and single top quark background samples are produced at NLO QCD using \POWHEG v2.0 and v1.0, respectively~\cite{Campbell:2014kua,Alioli:2009je,Re:2010bp}. Finally, the $\PW\PZ$ and $\PZ\PZ$ diboson productions are simulated at LO with \PYTHIA v8.205~\cite{Sjostrand:2014zea}, while the $\PV\gamma$ and $\PW\PW$ processes are simulated at NLO QCD using \MGvATNLO and \POWHEG~\cite{Melia:2011tj}, respectively.

In all cases, generated events are interfaced with \PYTHIA v8.205 or higher for the simulation of fragmentation, parton showering, and the underlying event description, using the parameters from the \textsc{cuetp8m1} tune~\cite{Khachatryan:2015pea}. In the case of LO (NLO) \MGvATNLO samples, partons from the matrix elements are matched to the parton shower description via the MLM~\cite{Alwall:2007fs} (FxFx~\cite{Frederix:2012ps}) scheme. The \NNPDF v3.0~\cite{Ball:2014uwa} parton distribution functions (PDFs) are used for all the matrix element calculations. Interactions of the final-state particles with the CMS detector are simulated with \GEANTfour~\cite{Agostinelli:2002hh}. Simulated events include the effects of pileup, and are weighted to reproduce the observed pileup distribution.

\section{Event selection}\label{sec:selection}

Events in the signal region (SR) are selected initially by the L1 trigger exploting the \ptmiss information, whose threshold varies between 60 and 90\GeV depending on the instantaneous luminosity. The \ptmiss at the L1 trigger is computed from the vector \pt sum of all the energy depositions in the calorimeters with ${\abs{\eta} < 3}$. Partial mistiming of signals in the forward region of the ECAL endcaps (${2.5  < \abs{\eta} < 3.0}$) led to a reduction in the L1 trigger efficiency. A correction for this effect was determined using an unbiased data sample. This correction was found to be about 1\% for \mjj of 200\GeV and it increases to about 20\% for \mjj larger than 3.5\TeV.

At the HLT level, events of interest are collected using triggers with thresholds of 110 or 120\GeV, depending on the data taking period, applied equally to both the missing transverse momentum computed at the trigger level (\mettrig) and the \mhttrig variable. The \mhttrig is defined as the magnitude of the vector \pt sum of the reconstructed jets at the trigger level in the event with  ${\pt > 20\GeV}$ and ${\abs{\eta} < 5}$. The energy fraction attributed to neutral hadrons in jets with ${\abs{\eta} < 3.0}$ is required to be less than 90\%, in order to remove spurious jets originating from detector noise. Both \mettrig and \mhttrig are calculated without including muon candidates, allowing the same triggers to be used also for selecting events in the muon CRs, which are used in the background estimation procedure described in Section~\ref{sec:background}.

Offline, events considered in the VBF search are required to have at least two jets with \pt larger than 80\,(40)\GeV for the leading (subleading) jet. Since the L1 trigger decision does not use information from the hadronic activity in the forward region, at least one of the two leading jets in the event is required to have ${\abs{\eta} < 3}$. To ensure a high and stable trigger efficiency, events are further required to have ${\ptmiss > 250\GeV}$.  The trigger efficiency is measured as a function of \mht, computed from jets with ${\pt > 30\GeV}$ and ${\abs{\eta} < 3}$. After correcting for the L1 mistiming inefficiency, these triggers are found to be fully efficient for events passing the analysis selection with ${\mht > 250\GeV}$. In addition, if the leading jet is within the geometrical acceptance of the tracker (${\abs{\eta} < 2.4}$), its energy fraction attributed to charged hadrons is required to be greater than 10\%, while the energy fraction attributed to neutral hadrons is required to be smaller than 80\%. These requirements, along with quality filters applied to tracks, muon candidates, and other physics objects, reduce the contamination arising from large misreconstructed \ptmiss from noncollision backgrounds~\cite{CMS-PAS-JME-16-004}.
To further suppress the contamination from QCD multijet events, in which a large \ptmiss may arise from a severe mismeasurement of the jet momentum, the jets in the event, with ${\pt > 30\GeV}$ and ${\abs{\eta} < 4.7}$, are required to not be aligned with the \ptvecmiss. The minimum value of the azimuthal angle between the \ptvecmiss vector and each jet ($\text{min}\dphijmet$) is required to be larger than 0.5\unit{rad}, where only the first four leading jets are included in the $\text{min}\dphijmet$ definition. This selection reduces the QCD multijet contamination to less than 1\% of the total background.

The two leading jets in VBF signal events typically show a large separation in $\eta$, large \mjj and a small azimuthal separation ($\abs{\dphijj}$).  The discriminatory power of $\abs{\dphijj}$ results from a combination of the spin-parity properties of the Higgs boson and the high-\pt regime explored by this search~\cite{Eboli:2000ze}, in which the two VBF jets tend to recoil against the invisible system. The \Zvvjets and \Wlvjets processes constitute the largest backgrounds in this search. The shape analysis primarily employs the large separation power of \mjj to discriminate between VBF signal and \Vjets backgrounds. Therefore, in this scenario, a set of loose requirements is applied on both \mjj and $\abs{\detajj}$, \ie $\abs{\detajj} > 1.0$ and $\mjj > 200\GeV$. To further reduce the \Vjets contamination, $\abs{\dphijj}$ is required to be smaller than 1.5\unit{rad} and the two jets must lie in opposite hemispheres, $\eta_{\mathrm{j1}}\,\eta_{\mathrm{j}2} < 0$. In the cut-and-count approach, \Zvvjets and \Wlvjets processes are suppressed by a more stringent event selection requiring ${\abs{\detajj}>4.0}$ and ${\mjj>1.3\TeV}$, while the requirement applied on $\abs{\dphijj}$ remains unchanged.

The \Wlvjets background is further suppressed by rejecting events that contain at least one isolated electron or muon with $\pt > 10\GeV$, or a \tauh candidate with $\pt > 18\GeV$ and $\abs{\eta} < 2.3$, where the isolation is required to be less than 25\,(16)\% of the muon (electron) \pt. With this strategy, prompt muons (electrons) are selected with an average efficiency of about 98\,(95)\%. In order to further reduce the contribution from \phojets and $\PV\gamma$ processes, events containing an isolated photon with $\pt > 15\GeV$ and $\abs{\eta} < 2.5$, passing identification criteria based on its ECAL shower shape~\cite{Khachatryan:2015iwa}, are vetoed.

Top quark backgrounds (\ttbar and single top quark processes) are suppressed by rejecting events in which at least one jet, with ${\pt > 20\GeV}$ and ${\abs{\eta} < 2.4}$, is identified as a \cPqb~quark jet using the combined secondary vertex (CSVv2) algorithm~\cite{Sirunyan:2017ezt}. A working point that yields a 60\% efficiency for tagging a \cPqb~quark jet and a 1\,(10)\% probability of misidentifying a light-flavor (\cPqc~quark) jet as a \cPqb~quark jet is used.

A summary of the selection criteria for the SR for both the shape and the cut-and-count analyses is shown in Table~\ref{tab:analysis_selection}.

\begin{table*}[htb]
    \centering
    \topcaption{Summary of the kinematic selections used to define the SR for both the shape and the cut-and-count analyses.}
    \cmsTable{
    \begin{tabular}{l c c c}
       \hline
       Observable                       & Shape analysis & Cut-and-count analysis & Target background \\
       \hline
       Leading (subleading) jet         & \multicolumn{2}{c}{$\pt>80\,(40)\GeV$, $\abs{\eta}<4.7$} & All \\
       \ptmiss                          & \multicolumn{2}{c}{$>$250\GeV}         & QCD multijet, \ttbar, \phojets, \Vjets  \\
       $\dphijmet$                      & \multicolumn{2}{c}{$>$0.5\unit{rad}}   & QCD multijet, \phojets \\
       Muons (electrons)                & \multicolumn{2}{c}{$\text{N}_{\mu,\Pe}=0$ with $\pt>10$\GeV, $\abs{\eta}<2.4\,(2.5)$} & \Wlvjets \\
       \tauh candidates                 & \multicolumn{2}{c}{$\text{N}_{\tauh}=0$ with $\pt>18$\GeV, $\abs{\eta}<2.3$}         & \Wlvjets \\
       Photons                          & \multicolumn{2}{c}{$\text{N}_{\gamma}=0$ with $\pt>15$\GeV, $\abs{\eta}<2.5$}        & \phojets, $\mathrm{V}\gamma$ \\
       \cPqb~quark jet                  & \multicolumn{2}{c}{$\text{N}_{\text{jet}}=0$ with $\pt>20$\GeV, $\text{CSVv2}>0.848$} & \ttbar, single top quark \\
       $\eta_{\mathrm{j1}}\,\eta_{\mathrm{j2}}$      & \multicolumn{2}{c}{$<$0} & \Zvvjets, \Wlvjets \\
       $\abs{\dphijj}$                  & \multicolumn{2}{c}{$<$1.5\unit{rad}}   & \Zvvjets, \Wlvjets \\
       $\abs{\detajj}$                  & $>$1 & $>$4             & \Zvvjets, \Wlvjets \\
       $\mjj$                           & $>$200\GeV & $>$1.3\TeV & \Zvvjets, \Wlvjets \\
       \hline
    \end{tabular}
     }
    \label{tab:analysis_selection}
\end{table*}

\section{Analysis strategy}\label{sec:background}

The search exploits the large \mjj and $\abs{\detajj}$ that characterize events from VBF Higgs boson production. In the shape analysis case, the signal is extracted by fitting the sum of the signal and background shapes to the binned \mjj distribution observed in data. The signal is expected to accumulate as an excess of events over the background at large values of \mjj. This strategy necessitates a precise estimation of the shape of the background \mjj distribution.

\begin{figure*}[htb]
\centering
\includegraphics[width=0.40\textwidth]{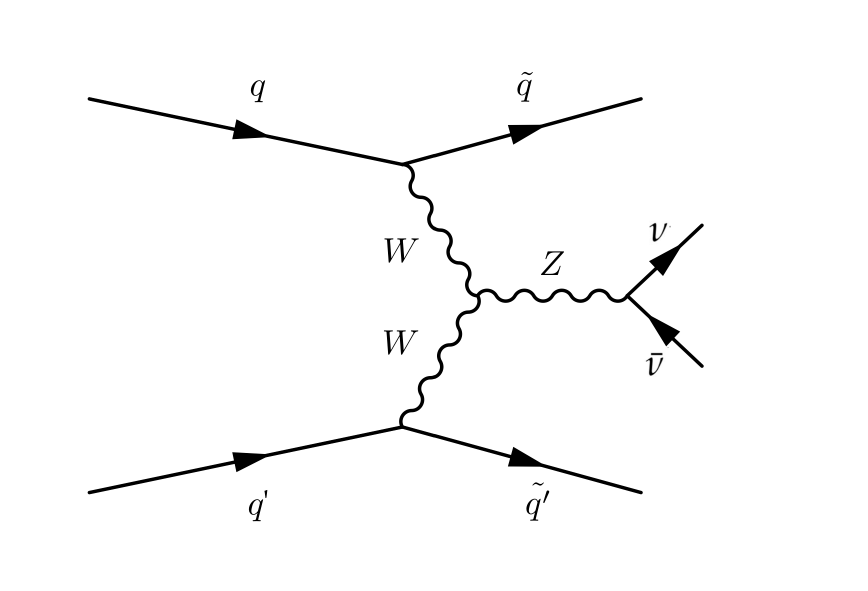}
\includegraphics[width=0.40\textwidth]{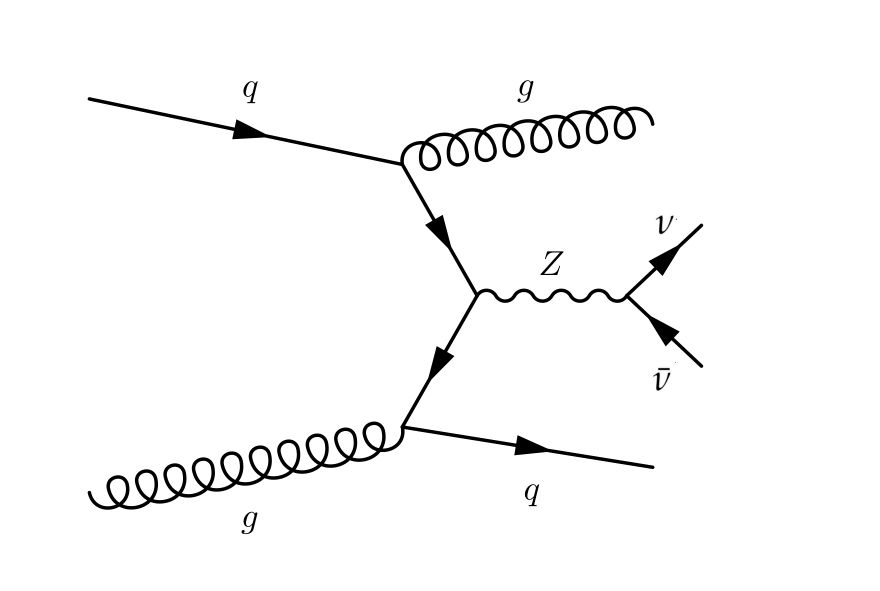}
\caption{Representative leading order Feynman diagrams for the production of a $\PZ$ boson in association with two partons arising from EW (left) and QCD (right) interactions. The left diagram contributes to the \Zvvjets (EW) production cross section, while the diagram on the right to the \Zvvjets (QCD) one. Diagrams for EW and QCD production of a $\PW$ boson in association with two jets are similar to those reported above for the \Zvvjets process.}
\label{fig:vjj_FD}
\end{figure*}

About 95\% of the total expected background in this search is due to the \Vjets processes, namely \Zvvjets and \Wlvjets. A fraction of the \Vjets background, referred to as \Vjets (EW), can be attributed to the EW production of a $\PZ$ or a $\PW$ boson in association with two jets. A representative Feynman diagram contributing to \PV{}+jets (EW) production is shown in Fig.~\ref{fig:vjj_FD} (left). The remaining \Vjets contribution arises from the production of a vector boson in association with QCD radiation, as shown in Fig.~\ref{fig:vjj_FD} (right). This is referred to as the \Vjets (QCD) background. For both EW and QCD productions, the expected \Zvvjets rate in the SR is about two times larger than the \Wlvjets contribution.

\begin{figure*}[htb]
\centering
\includegraphics[width=0.325\textwidth]{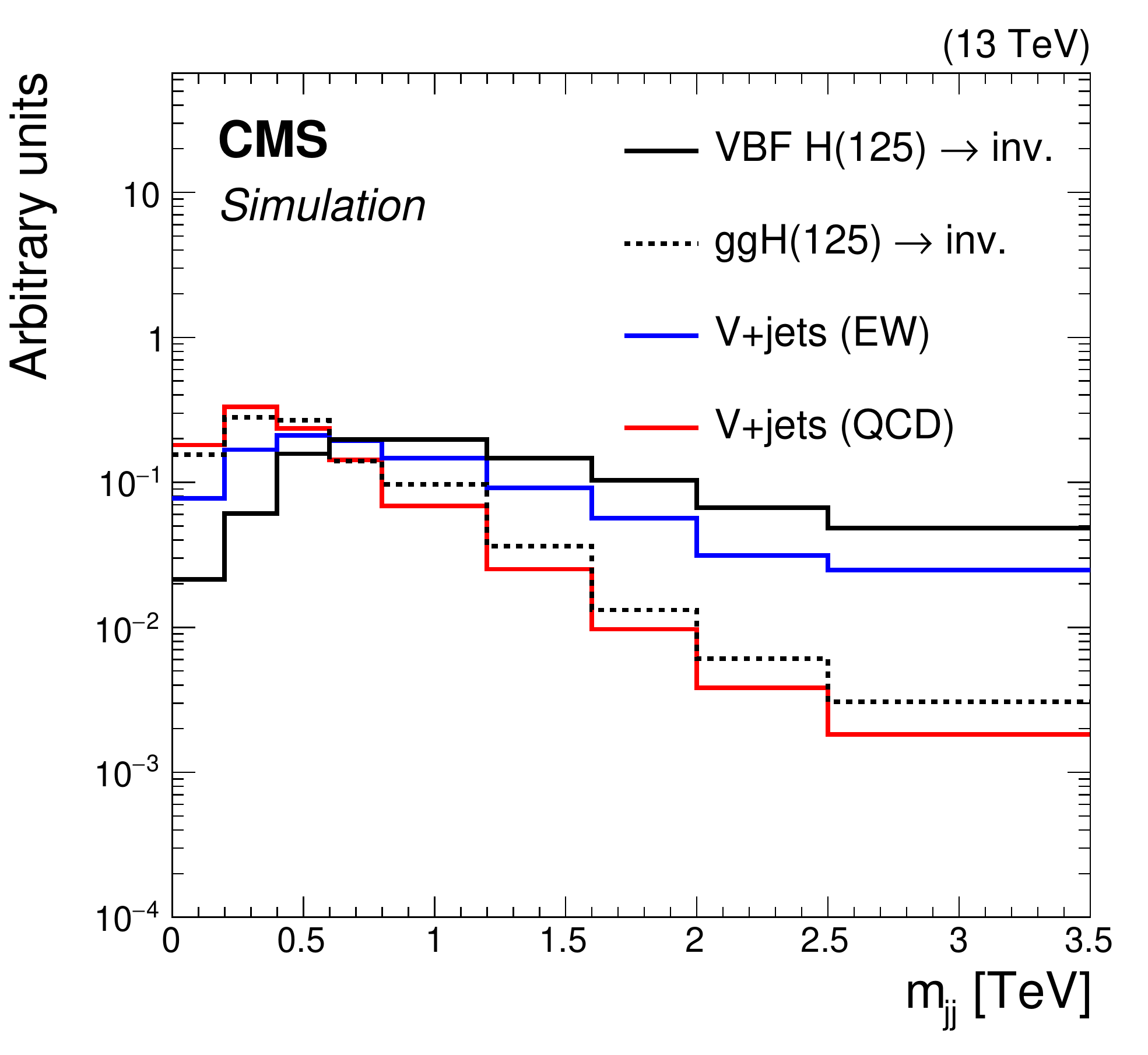}
\includegraphics[width=0.325\textwidth]{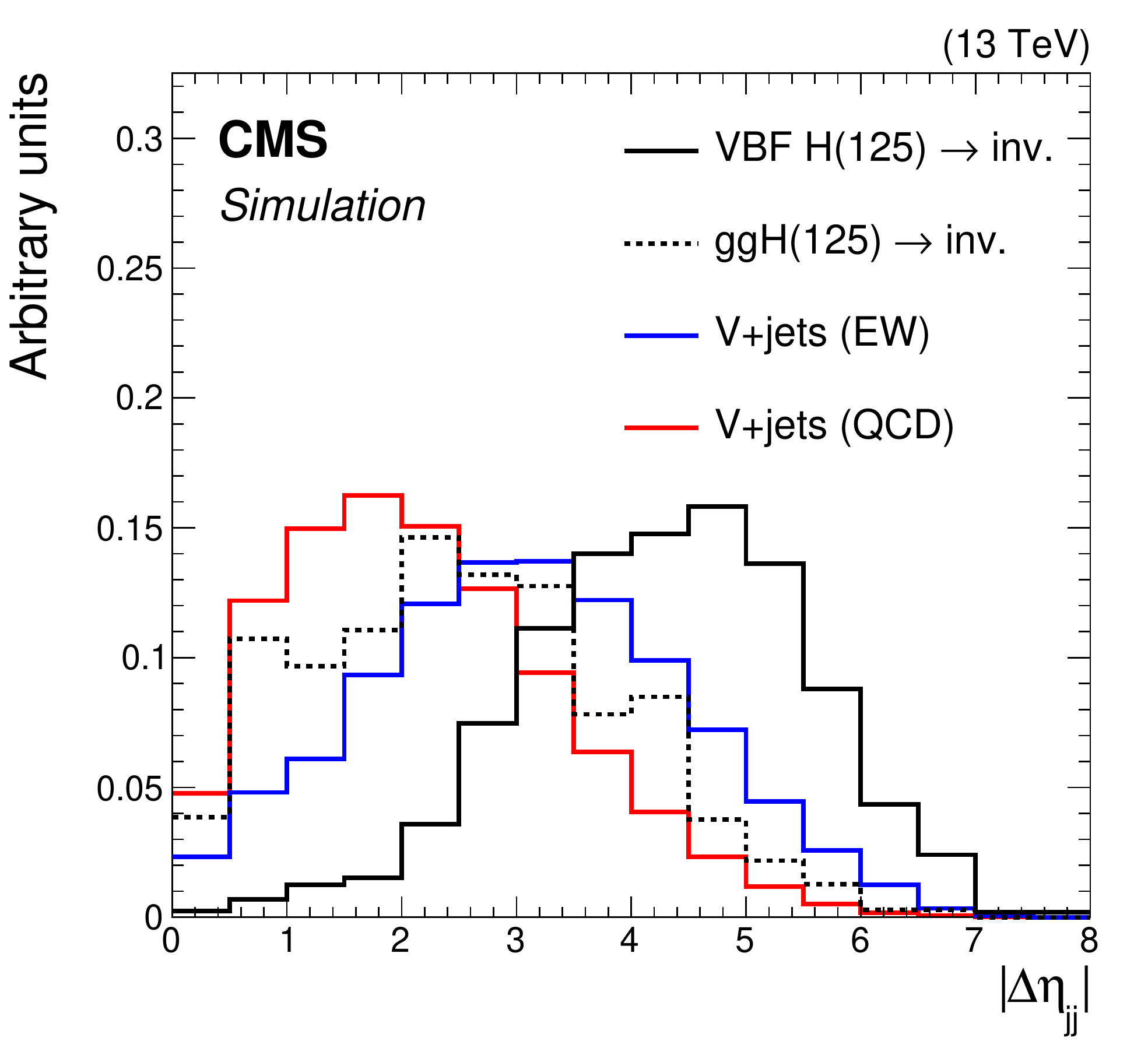}
\includegraphics[width=0.325\textwidth]{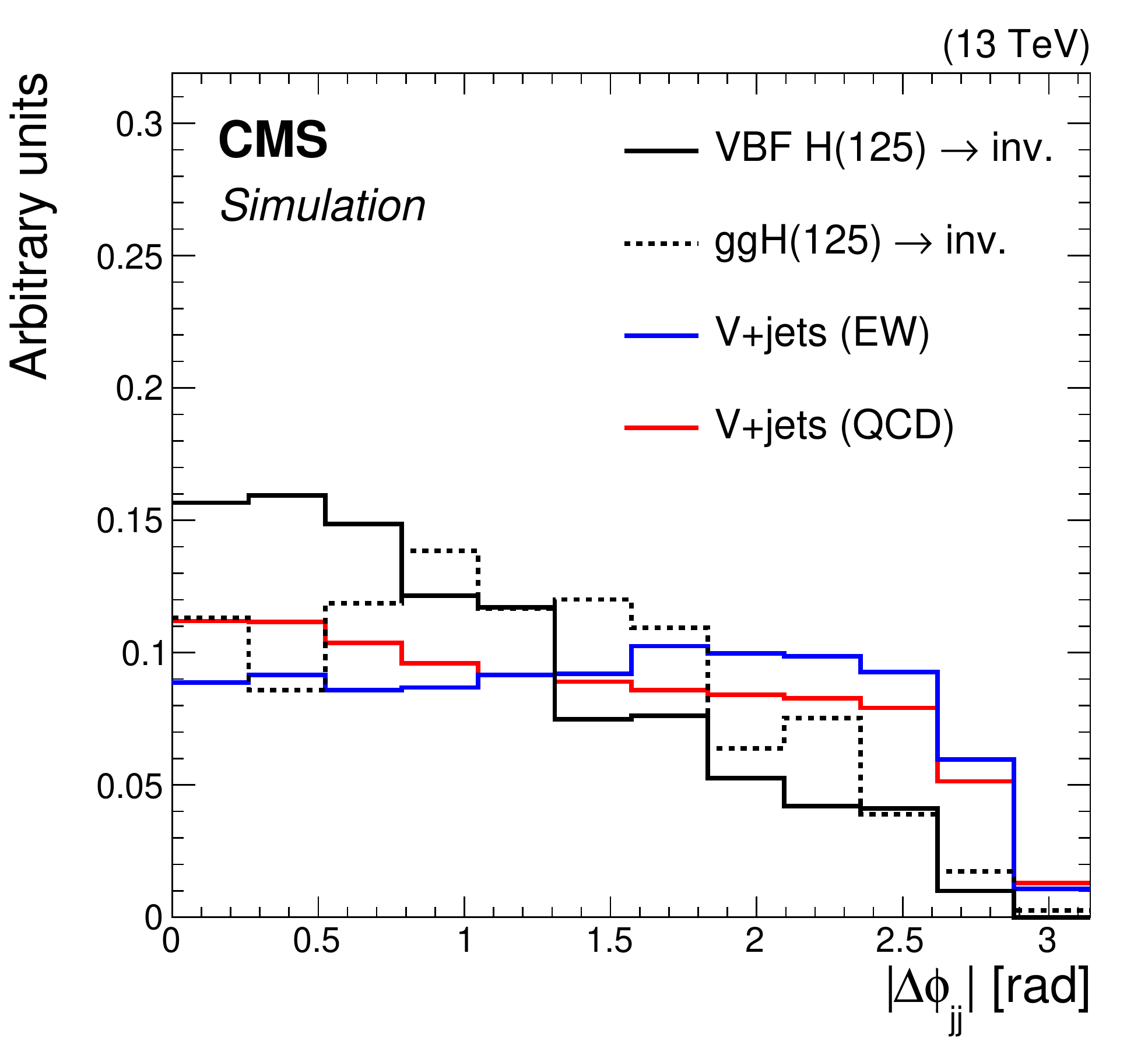}
\caption{Comparison between the shapes of the \mjj (left), $\abs{\detajj}$ (middle) and $\abs{\dphijj}$ (right) distributions of signal events, produced by VBF (solid black) and $\Pg\Pg\PH$ (dashed black) mechanisms, and \Vjets backgrounds from both QCD (solid red) and EW (solid blue) production. Both signal and background distributions are scaled in order to have unit area. Distributions are obtained from simulated events passed through the CMS event reconstruction.}
\label{fig:signal_vs_background_shapes}
\end{figure*}

A comparison of the shapes of the key discriminating observables used in this analysis, obtained after applying the requirements listed in Table~\ref{tab:analysis_selection} except for those on \mjj, $\abs{\detajj}$ and $\abs{\dphijj}$, is shown in Fig.~\ref{fig:signal_vs_background_shapes} for simulated signal and \Vjets background events. From these distributions, it can be seen that the \Vjets (EW) background is kinematically similar to the VBF Higgs boson signal. Therefore, its contribution to the total \Vjets background rate increases when the two leading jets have large \mjj and $\abs{\detajj}$. The \Vjets (EW) process constitutes about $2\%$ of the total \Vjets background for \mjj around 200\GeV. Its contribution increases to about 20\% for $\mjj \approx 1.5\TeV$, and is more than 50\% for $\mjj > 3\TeV$.

\subsection{Overview of the \texorpdfstring{\PV{}+jets}{V+jets} background estimation}

The \Zjets and \Wjets backgrounds are estimated using four mutually exclusive CRs. These include a dimuon and a dielectron CR consisting mostly of \Zlljets events that are kinematically similar to \Zvvjets background if the presence of the two leptons in the event is ignored. The \Wjets background is estimated using CRs consisting of single-muon and single-electron events stemming mainly from leptonic decays of a $\PW$ boson. In contrast to the \Wjets background in the SR, the single-lepton CRs consist of leptons that fall within the detector acceptance and pass the identification requirements. The \ptmiss in all the CRs is calculated by excluding the contribution of the identified leptons. Therefore, it corresponds to the \pt of the hadronic recoil system, which resembles the \ptmiss expected from the \Vjets backgrounds in the SR.

The event yield in the dilepton CRs is considerably smaller than the \Zvvjets contribution in the SR because the \Zll branching fraction, where $\ell = \mu$ or \Pe, is six times smaller than the \Zvv branching fraction. Consequently, the dilepton CRs have a limited statistical power to constrain the \Zvvjets background by themselves. In contrast, the yield of the single-lepton CRs is comparable to the \Zvvjets background. Furthermore, the \Zvvjets and \Wlvjets processes are kinematically similar if the presence of the charged lepton is ignored. The theoretical uncertainties involved in the prediction of the \Zjets and \Wjets cross sections largely cancel out in their ratio. Therefore, this ratio is predicted very reliably by the simulation and can be used as a constraint to connect the statistically rich single-lepton CRs to the \Zvvjets background in the SR.

The predictions for the \Vjets processes obtained from simulation are referred to as ``pre-fit'' expectations, and are considered to be the initial estimates for the \Vjets yields in the CRs and SR. These \Vjets yields are then treated as freely floating parameters, and are fit to the data in all CRs and the SR. The \Vjets yields obtained from this fit are referred to as ``post-fit'' estimates, and serve as the final \Vjets background predictions in the analysis.

\subsection{Definition of control regions}

Dimuon and single-muon CRs are selected using the same L1 and HLT \ptmiss-based triggers that are used to collect events in the SR. Dimuon events are required to contain exactly two oppositely charged muons with $\pt > 10\GeV$ that form an invariant mass ($m_{\mu\mu}$) between 60 and 120\GeV, which is compatible with a $\PZ$ boson decay. Events with additional electrons or photons are rejected. At least one of the two muons must have $\pt>20\GeV$, and is required to pass tighter identification criteria based on the number of measurements in the tracker and the muon systems, the quality of the muon track fit, and the consistency of the muon track with the primary vertex. The isolation, as defined in Section~\ref{sec:reconstruction}, is required to be smaller than 15\% of the muon \pt. These tightly identified muons are selected with an average efficiency of 90\%.

In the single-muon CR, events are required to contain exactly one muon with ${\pt > 20\GeV}$, passing both tight identification and isolation requirements. The transverse mass (\mT) of the muon-\ptmiss system is computed as ${\mT = \sqrt{\smash[b]{2\ptmiss \pt^{\mu} (1-\cos\Delta\phi)}}}$, where $\pt^{\mu}$ is the \pt of the muon, and $\Delta\phi$ is the angle between $\ptvec^{\mu}$ and \ptvecmiss in the transverse plane. The \mT is required to be smaller than 160\GeV, and no additional electrons or photons are allowed in the event.

Events in the dielectron and single-electron CRs are collected mainly using a single-electron trigger with a \pt threshold of 27\GeV. In the case of dielectron events where the $\PZ$ boson has $\pt > 600 \GeV$, the two electrons have a small angular separation, and are likely to get included in each other's isolation cones.
This results in an inefficiency for the chosen trigger that imposes isolation requirements on electron candidates. This inefficiency is mitigated by including events collected by a single-electron trigger with a \pt threshold of 105\GeV and no isolation requirements on the electron candidate.

The dielectron events are required to contain exactly two oppositely charged electrons with $\pt > 10\GeV$ and no additional muons or photons. As in the case of the dimuon events, the invariant mass of the dielectron system is required to be between 60 and 120\GeV. At least one of the two electrons must have ${\pt > 40\GeV}$, and is required to pass a tight identification criterion based on the shower shape of its ECAL energy deposit, the matching of the electron track to the ECAL energy cluster, and the consistency of the electron track with the primary vertex. Furthermore, the isolation is required to be smaller than 6\% of the electron \pt. These selection requirements for electrons have an average efficiency of 70\%.

Events in the single-electron CR are required to contain exactly one tightly identified and isolated electron with $\pt > 40\GeV$; no additional muons or photons are allowed. The contamination from QCD multijet events is reduced by requiring $\ptmiss > 60\GeV$ and $\mT < 160 \GeV$.

Events in the CRs must also satisfy the requirements imposed on events in the SR. When doing so, the negative \pt of the hadronic recoil system is used instead of the \ptmiss in the event.

\subsection{Estimation of \texorpdfstring{\PV{}+jets}{V+jets} backgrounds}\label{sec:vjets_background_estimation}

The \Vjets yields in the CRs are translated to the background estimates in the SR using transfer factors that are derived from simulation. The transfer factors are defined as the ratio of the yields of a given \Vjets background in the SR and the corresponding process measured in each CR.

The transfer factors for the dilepton CRs account for the difference in the branching fractions of the \Zvv and \Zll decays, and the $\gamma^{*}(\ell\ell)$ contribution, as well as the impact of lepton acceptance and selection efficiencies. In the case of dielectron events, the transfer factors also account for the difference in trigger efficiencies. Transfer factors between the \Wlvjets event yields in the single-lepton CRs and the \Wjets background estimate in the SR take into account the effect of lepton acceptance, selection efficiencies, and lepton and \tauh veto efficiencies, as well as the difference in trigger efficiencies in the case of the single-electron CR.

The constraint on the ratio of the cross sections of the \Zjets and \Wjets processes, which is used to connect the single-lepton CRs to the \Zvvjets in the SR, is also implemented as a transfer factor, and is computed as the ratio of the \Zvvjets and \Wlvjets yields in the SR. In order to have the most precise estimate of this constraint, the LO simulations for the \Zjets (QCD) and the \Wjets (QCD) processes are corrected using boson \pt and ${\mjj\text{--\,dependent}}$ NLO QCD $K$-factors derived with \MGvATNLO. The \Zjets and \Wjets simulations are also corrected as a function of boson \pt with NLO EW $K$-factors derived from theoretical calculations~\cite{Lindert:2017olm}. Similarly, \Zjets (EW) and \Wjets (EW) processes are corrected with NLO QCD $K$-factors derived
using the \textsc{vbfnlo} event generator~\cite{Arnold:2008rz,Baglio:2014uba} as a function of boson \pt and \mjj.

The \Vjets background yields are determined using a maximum-likelihood fit, performed simultaneously across all CRs and the SR. The likelihood function is defined as:
\begin{linenomath}
\ifthenelse{\boolean{cms@external}}{
\begin{equation}
\begin{aligned}
\mathcal{L}&(\mu,\kappaz, \boldsymbol{\theta}) =  \\
&\prod_{i} \mathrm{P}\Biggl(d_{i} \Big{|} B_{i}(\boldsymbol{\theta}) + (1+f_{i}(\boldsymbol{\theta})_{\mathrm{Q}}) \kappazi \Biggr.\\
&\qquad\Biggl.+ R^{\PZ}_{i} (1+f_{i}(\boldsymbol{\theta})_{\mathrm{E}}) \kappazi + \mu S_{i}(\boldsymbol{\theta})\Biggr) \\
&\prod_{i} \mathrm{P} \left(d^{\mu\mu}_{i} \Big{|} B^{\mu\mu}_{i}(\boldsymbol{\theta}) +\frac{\kappazi }{R^{\mu\mu}_{i} (\boldsymbol{\theta})_{\mathrm{Q}}} + \frac{R^{\PZ}_{i} \kappazi }{R^{\mu\mu}_{i} (\boldsymbol{\theta})_{\mathrm{E}}} \right) \\
& \prod_{i} \mathrm{P}\left(d^{\Pe\Pe}_{i} \Big{|} B^{\Pe\Pe}_{i}(\boldsymbol{\theta}) +\frac{\kappazi }{R^{\Pe\Pe}_{i} (\boldsymbol{\theta})_{\mathrm{Q}}}+\frac{R^{\PZ}_{i}  \kappazi }{R^{\Pe\Pe}_{i} (\boldsymbol{\theta})_{\mathrm{E}}} \right) \\
& \prod_{i} \mathrm{P}\left(d^{\mu}_{i} \Big{|} B^{\mu}_{i}(\boldsymbol{\theta}) +\frac{f_{i}(\boldsymbol{\theta})_{\mathrm{Q}}\,\kappazi}{R^{\mu}_{i}(\boldsymbol{\theta})_{\mathrm{Q}}}+\frac{R^{\PZ}_{i} f_{i}(\boldsymbol{\theta})_{\mathrm{E}}\,\kappazi }{R^{\mu}_{i} (\boldsymbol{\theta})_{\mathrm{E}}} \right)  \\
& \prod_{i} \mathrm{P}\left(d^{\Pe}_{i} \Big{|} B^{\Pe}_{i}(\boldsymbol{\theta}) +\frac{f_{i}(\boldsymbol{\theta})_{\mathrm{Q}}\,\kappazi}{R^{\Pe}_{i}(\boldsymbol{\theta})_{\mathrm{Q}}}+\frac{R^{\PZ}_{i}  f_{i}(\boldsymbol{\theta})_{\mathrm{E}}\,\kappazi }{R^{\Pe}_{i} (\boldsymbol{\theta})_{\mathrm{E}}} \right) \, \prod_{j} \mathrm{P}(\theta)
\end{aligned}
\end{equation}
}{
\begin{equation}
\begin{aligned}
\mathcal{L} (\mu,\kappaz, \boldsymbol{\theta}) = & \prod_{i} \mathrm{P}\left(d_{i} \Big{|} B_{i}(\boldsymbol{\theta}) + (1+f_{i}(\boldsymbol{\theta})_{\mathrm{Q}}) \kappazi + R^{\PZ}_{i} (1+f_{i}(\boldsymbol{\theta})_{\mathrm{E}}) \kappazi + \mu S_{i}(\boldsymbol{\theta})\right) \\
&\prod_{i} \mathrm{P} \left(d^{\mu\mu}_{i} \Big{|} B^{\mu\mu}_{i}(\boldsymbol{\theta}) +\frac{\kappazi }{R^{\mu\mu}_{i} (\boldsymbol{\theta})_{\mathrm{Q}}} + \frac{R^{\PZ}_{i} \kappazi }{R^{\mu\mu}_{i} (\boldsymbol{\theta})_{\mathrm{E}}} \right) \\
& \prod_{i} \mathrm{P}\left(d^{\Pe\Pe}_{i} \Big{|} B^{\Pe\Pe}_{i}(\boldsymbol{\theta}) +\frac{\kappazi }{R^{\Pe\Pe}_{i} (\boldsymbol{\theta})_{\mathrm{Q}}}+\frac{R^{\PZ}_{i}  \kappazi }{R^{\Pe\Pe}_{i} (\boldsymbol{\theta})_{\mathrm{E}}} \right) \\
& \prod_{i} \mathrm{P}\left(d^{\mu}_{i} \Big{|} B^{\mu}_{i}(\boldsymbol{\theta}) +\frac{f_{i}(\boldsymbol{\theta})_{\mathrm{Q}}\,\kappazi}{R^{\mu}_{i}(\boldsymbol{\theta})_{\mathrm{Q}}}+\frac{R^{\PZ}_{i} f_{i}(\boldsymbol{\theta})_{\mathrm{E}}\,\kappazi }{R^{\mu}_{i} (\boldsymbol{\theta})_{\mathrm{E}}} \right)  \\
& \prod_{i} \mathrm{P}\left(d^{\Pe}_{i} \Big{|} B^{\Pe}_{i}(\boldsymbol{\theta}) +\frac{f_{i}(\boldsymbol{\theta})_{\mathrm{Q}}\,\kappazi}{R^{\Pe}_{i}(\boldsymbol{\theta})_{\mathrm{Q}}}+\frac{R^{\PZ}_{i}  f_{i}(\boldsymbol{\theta})_{\mathrm{E}}\,\kappazi }{R^{\Pe}_{i} (\boldsymbol{\theta})_{\mathrm{E}}} \right) \, \prod_{j} \mathrm{P}(\theta_{j})
\end{aligned}
\end{equation}
}
\end{linenomath}

where ${\mathrm{P}(x|y) = y^{x}\re^{-y}/x!}$. The symbol $i$ denotes each bin of the \mjj distribution in the shape analysis, while, in the cut-and-count case, $i$ stands for a single bin that represents the event yields obtained at the end of the event selection. The symbols $d^{\mu\mu}_{i}$, $d^{\Pe\Pe}_{i}$, $d^{\mu}_{i}$, $d^{\Pe}_{i}$, and $d_{i}$ denote the observed number of events in each bin $i$ of the dimuon, dielectron, single-muon, single-electron CRs, and the SR, respectively. The symbols $f_{i}(\boldsymbol{\theta})_{\mathrm{Q}}$ and $f_{i}(\boldsymbol{\theta})_{\mathrm{E}}$ indicate the ratios between the \Wlvjets and \Zvvjets backgrounds in the SR from QCD and EW production, respectively.
The symbols $R^{\mu\mu}_{i}(\boldsymbol{\theta})_{\mathrm{Q}}$, $R^{\Pe\Pe}_{i}(\boldsymbol{\theta})_{\mathrm{Q}}$, $R^{\mu}_{i}(\boldsymbol{\theta})_{\mathrm{Q}}$, and $R^{\Pe}_{i}(\boldsymbol{\theta})_{\mathrm{Q}}$ are the transfer factors relating the dimuon, dielectron, single-muon, and single-electron CRs, respectively, to the SR for the \Vjets (QCD) processes.
Similarly, $R^{\mu\mu}_{i}(\boldsymbol{\theta})_{\mathrm{E}}$, $R^{\Pe\Pe}_{i}(\boldsymbol{\theta})_{\mathrm{E}}$, $R^{\mu}_{i}(\boldsymbol{\theta})_{\mathrm{E}}$, and $R^{\Pe}_{i}(\boldsymbol{\theta})_{\mathrm{E}}$ indicate the transfer factors for the \Vjets (EW) processes. The parameters $\kappazi$ represent the yield of the \Zvvjets (QCD) background in each bin $i$ of the SR, and are left to float freely in the fit. In a given bin, the \Zvvjets (EW) background yield is obtained from $\kappazi$ through the transfer factor $R_{i}^{\PZ}$ that represents the ratio between the \Zvvjets (QCD) and \Zvvjets (EW) processes. The contributions from subleading backgrounds in each region are estimated directly from simulation and they are denoted by $B^{\mu\mu}_{i}$, $B^{\Pe\Pe}_{i}$, $B^{\mu}_{i}$, $B^{\Pe}_{i}$ and $B_{i}$. Finally, the likelihood also includes a signal term in which $S_i$ represents the expected signal prediction, while $\mu = \sigmabr$ denotes the signal strength parameter.

Systematic uncertainties are modeled as constrained nuisance parameters ($\boldsymbol{\theta}$), for which log-normal or Gaussian priors, indicated by $\mathrm{P}_{j}(\theta)$ in previous equation, are considered. The systematic uncertainties in the \Vjets background estimates are introduced in the likelihood as variations of the transfer factors. These include theoretical uncertainties in the \Zjets to \Wjets differential cross section ratio for both the QCD and EW processes due to the choice of the renormalization and the factorization scales, as well as the choice of the PDFs. The QCD scale variations are assumed to be uncorrelated between the \Zjets and \Wjets processes, and therefore they do not cancel in the \Zjets to \Wjets cross section ratio. This results in larger uncertainties compared to those from NLO calculations recommended in Ref.~\cite{Lindert:2017olm}. The uncertainty due to the choice of the renormalization scale varies between 8 and 12\% as a function of \mjj for both \Zjets/\Wjets (QCD) and (EW) ratios. Similarly, the uncertainty due to the choice of the factorization scale varies between 2 and 7\%. This also covers the uncertainty in the \Zjets/\Wjets cross section ratio due to the interference between the \Vjets (QCD) and \Vjets (EW) processes, which is not included in the simulation. The PDF uncertainties are assumed to be correlated across \Vjets processes, resulting in a residual uncertainty smaller than 1\% on the \Zjets/\Wjets cross section ratio. The uncertainties related to NLO EW corrections to the \Vjets (QCD) processes are estimated according to the recommendations in Ref.~\cite{Lindert:2017olm}, and are found to be about 1--2\% across the entire \mjj spectrum. Additional uncertainties included in the transfer factors include uncertainties in the reconstruction efficiencies of leptons (around 1\% per muon or electron), the selection efficiencies of leptons (about 1\% per muon, 1.5\% per electron), the veto efficiency of leptons (around 2\% per muon, 1\% per electron) and \tauh candidates (about 3.5\% per \tauh), the knowledge of the jet energy scale (1--2\%), and the efficiency of the electron (around 1\%) and \ptmiss triggers (about 2\%).

The full set of systematic uncertainties related to the \Vjets transfer factors are listed in Table~\ref{tab:vbfsysunc}. Before any fit is performed, the total uncertainty in the expected background in the SR ranges between 4.5 and 6\% as a function of \mjj, dominated by the theoretical uncertainties in the \Zjets to \Wjets cross section ratio for both QCD and EW production. The impact of each source of systematic uncertainty, as reported in Table~\ref{tab:vbfsysunc} in the context of the shape analysis, is defined as the maximum difference in the fitted value of the signal strength, ${\sigmabr}$, obtained by varying the associated nuisance parameter within one standard deviation of its maximum likelihood estimate. In this procedure, the per-bin $\kappazi$ parameters are profiled when a given nuisance parameter is shifted from its best fit estimate.

\begin{table*}[htb]
    \centering
    \topcaption{Experimental and theoretical sources of systematic uncertainties on the \Vjets transfer factors, which enter in the simultaneous fit, used to estimate the \Vjets backgrounds, as constrained nuisance parameters. In addition, the impact on the fitted signal strength, ${\sigmabr}$, is reported in the last column estimated after performing the \mjj shape fit to the observed data across signal and control regions.}
    \cmsTable{
    \begin{tabular}{l l c c}
       \hline
       Source of uncertainty & Ratios & Uncertainty vs. \mjj & Impact on $\brhinv$\\ [\cmsTabSkip]
       \hline
       \multicolumn{4}{c}{Theoretical uncertainties} \\ [\cmsTabSkip]
       Ren. scale \Vjets (EW)   & \Zvv/\Wlv (EW)  & 9--12\%  & 48\% \\
       Ren. scale \Vjets (QCD)  & \Zvv/\Wlv (QCD) & 9--12\%  & 25\% \\
       Fac. scale \Vjets (EW)   & \Zvv/\Wlv (EW)  & 2--7\%   & 4\% \\
       Fac. scale \Vjets (QCD)  & \Zvv/\Wlv (QCD) & 2--7\%   & 2\% \\
       PDF \Vjets (QCD)         & \Zvv/\Wlv (QCD) & 0.5--1\% & $<$1\%\\
       PDF \Vjets (EW)          & \Zvv/\Wlv (EW)  & 0.5--1\% & $<$1\%\\
       NLO EW corr.             & \Zvv/\Wlv (QCD) & 1--2\%   & $<$1\%\\[\cmsTabSkip]
       \multicolumn{4}{c}{Experimental uncertainties}  \\ [\cmsTabSkip]
       Muon reco. eff.           & \Zmm/\Zvv, \Wmn/\Wlv & $\approx$1\%(per lepton) & 8\%\\
       Electron reco. eff.       & \Zee/\Zvv, \Wen/\Wlv & $\approx$1\%(per lepton) & 3\%\\
       Muon id. eff.             & \Zmm/\Zvv, \Wmn/\Wlv & $\approx$1\%(per lepton) & 8\%\\
       Electron id. eff.         & \Zee/\Zvv, \Wen/\Wlv & $\approx$1.5\%(per lepton) & 4\%\\
       Muon veto                 & \Zvv/\Wlv, \WCR/\Wlv & $\approx$2.5\,(2)\% for EW (QCD) & 7\%\\
       Electron veto             & \Zvv/\Wlv, \WCR/\Wlv & $\approx$1.5\,(1)\% for EW (QCD) & 5\%\\
       $\tau$ veto               & \Zvv/\Wlv, \WCR/\Wlv & $\approx$3.5\,(3)\% for EW (QCD) & 13\%\\
       Jet energy scale          & \ZCR/\Zvv, \WCR/\Wlv & $\approx$1\,(2)\% for \PZ/\PZ (\PW/\PW) & 4\%\\
       Electron trigger          & \Zee/\Zvv, \Wen/\Wlv & $\approx$1\% & $<$1\% \\
       \ptmiss trigger              & All ratios        & $\approx$2\% & 18\%\\
       \hline
    \end{tabular}
}
    \label{tab:vbfsysunc}
\end{table*}

\begin{figure*}[!htb]
\centering
\includegraphics[width=0.44\textwidth]{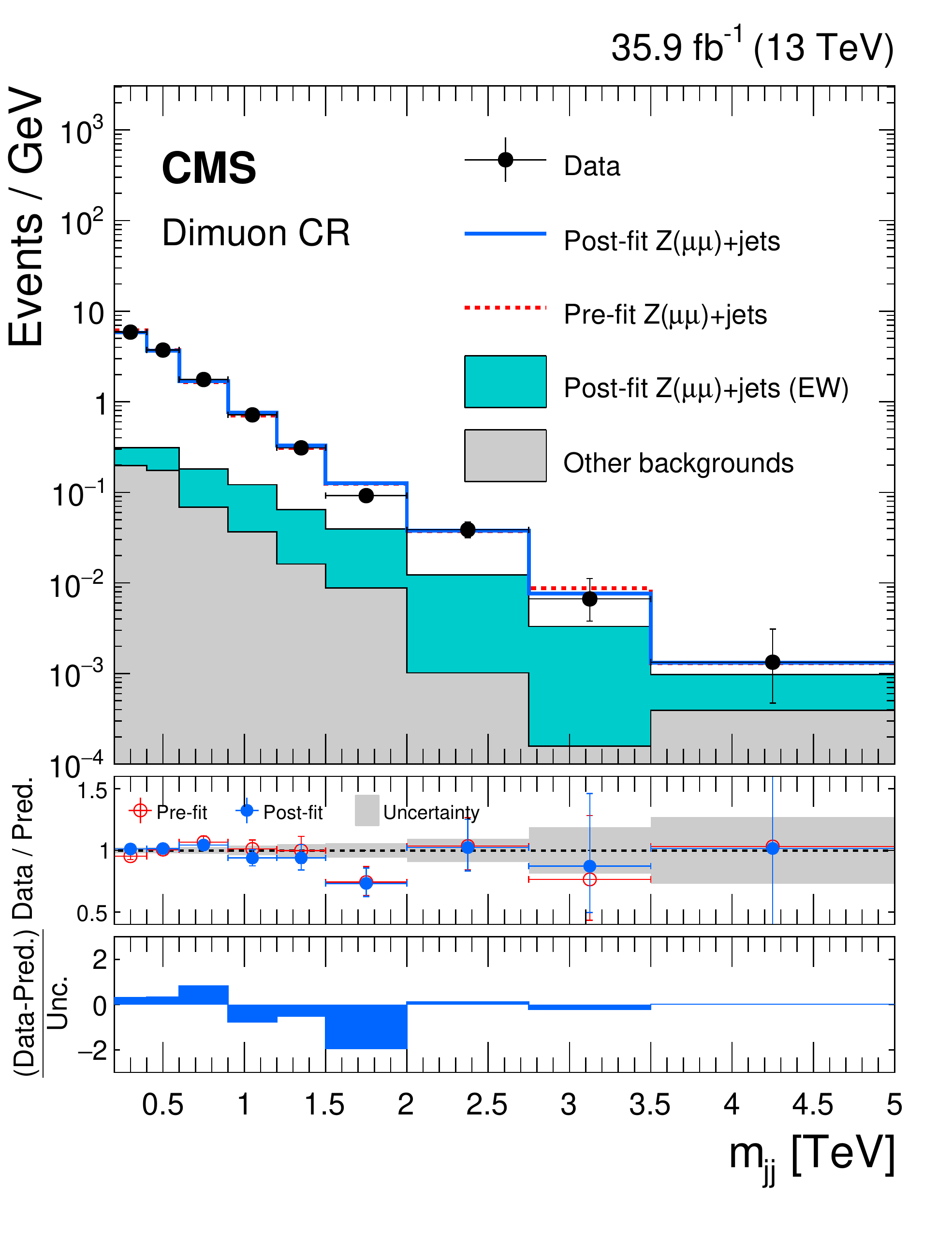}
\includegraphics[width=0.44\textwidth]{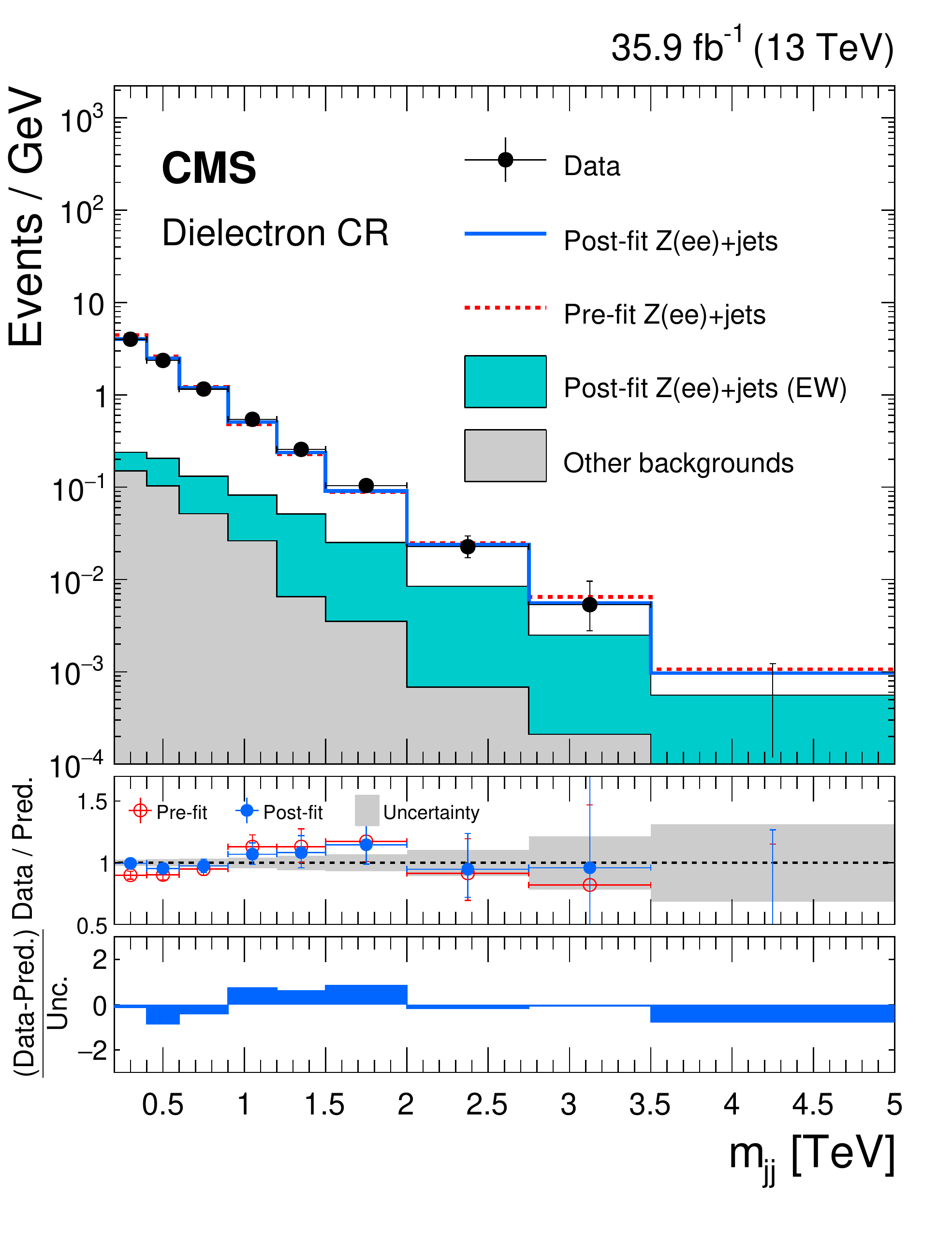}
\includegraphics[width=0.44\textwidth]{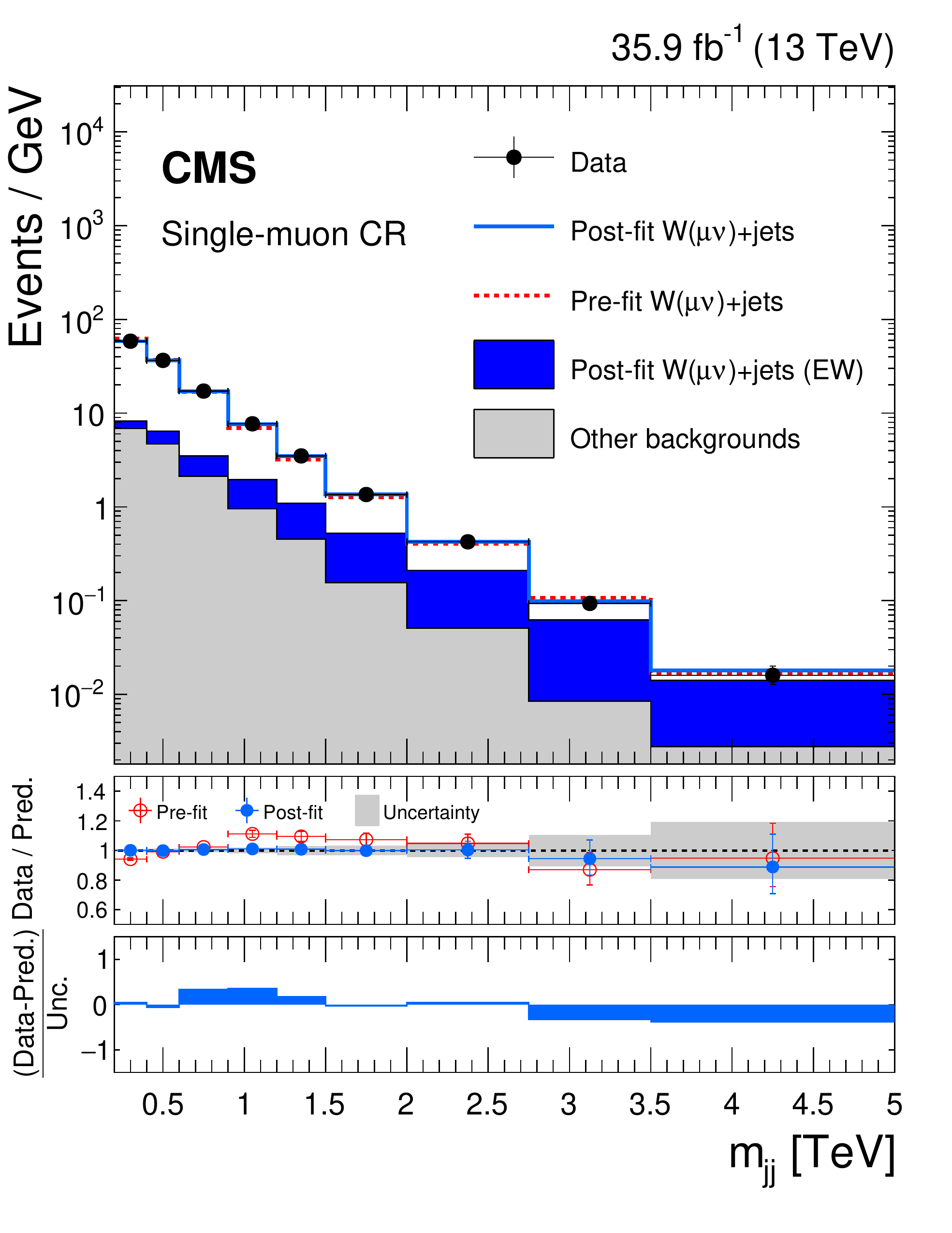}
\includegraphics[width=0.44\textwidth]{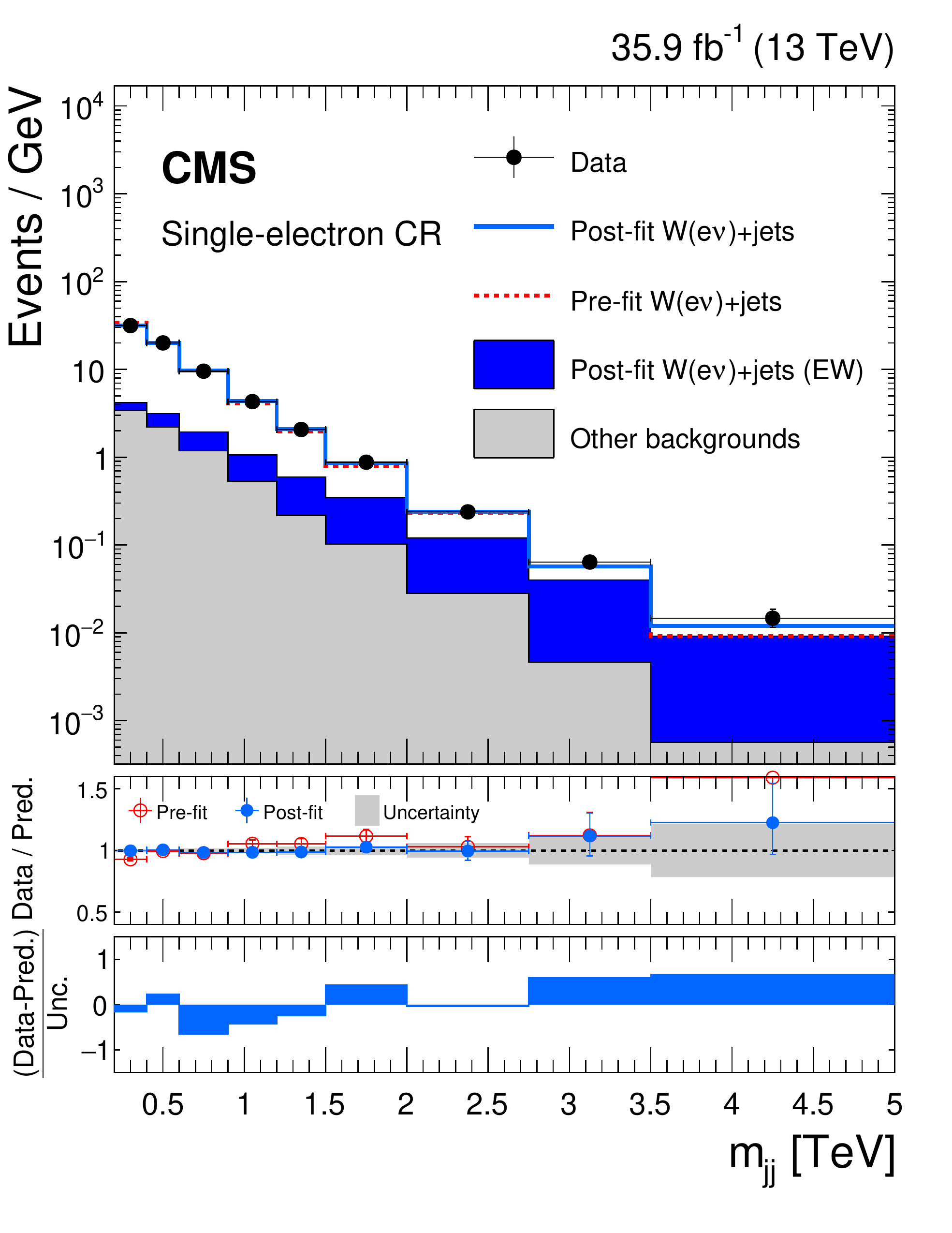}
\caption{
The \mjj distributions in the dimuon (top left), dielectron (top right), single-muon (bottom left), and single-electron (bottom right) CRs as computed in the shape analysis. Prediction from simulation (pre-fit estimate) is shown by the dashed red line. The solid blue line shows the \Vjets expectation after fitting the data in all the CRs. The filled histograms indicate all processes other than \Vjets (QCD). The last bin includes all events with ${\mjj > 3.5\TeV}$. Ratios of data and the pre-fit background (red points) and the post-fit background prediction (blue points) are shown. The gray band in the ratio panel indicates the total uncertainty after performing the fit. The lowest panel shows the difference between data and the post-fit background estimate relative to the post-fit background uncertainty.}
\label{fig:fit_controlregion}
\end{figure*}

The \mjj distributions in the dilepton and single-lepton CRs are shown in Fig.~\ref{fig:fit_controlregion}. The pre-fit predictions for the \Vjets processes are shown in red. These indicate the level of agreement between data and simulation before a fit is performed. An estimate of the \Vjets backgrounds is then obtained by fitting the data across all the CRs. This is depicted by the blue line in Fig.~\ref{fig:fit_controlregion}. This fit is referred to as the ``CR-only'' fit, since it does not impose any constraint on the \Vjets yields due to the data in the SR.

\begin{figure*}[htb]
\centering
\includegraphics[width=0.48\textwidth]{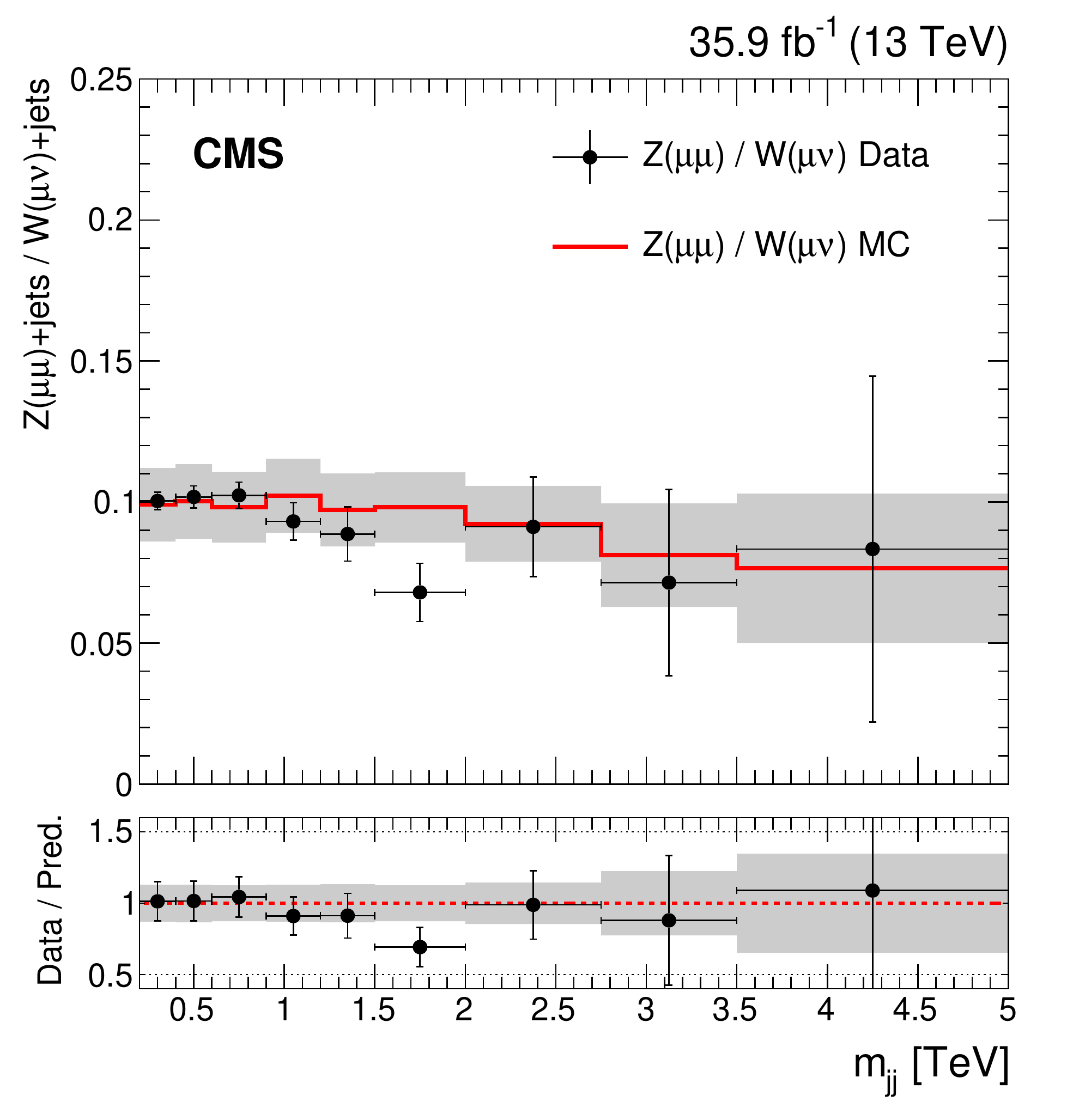}
\includegraphics[width=0.48\textwidth]{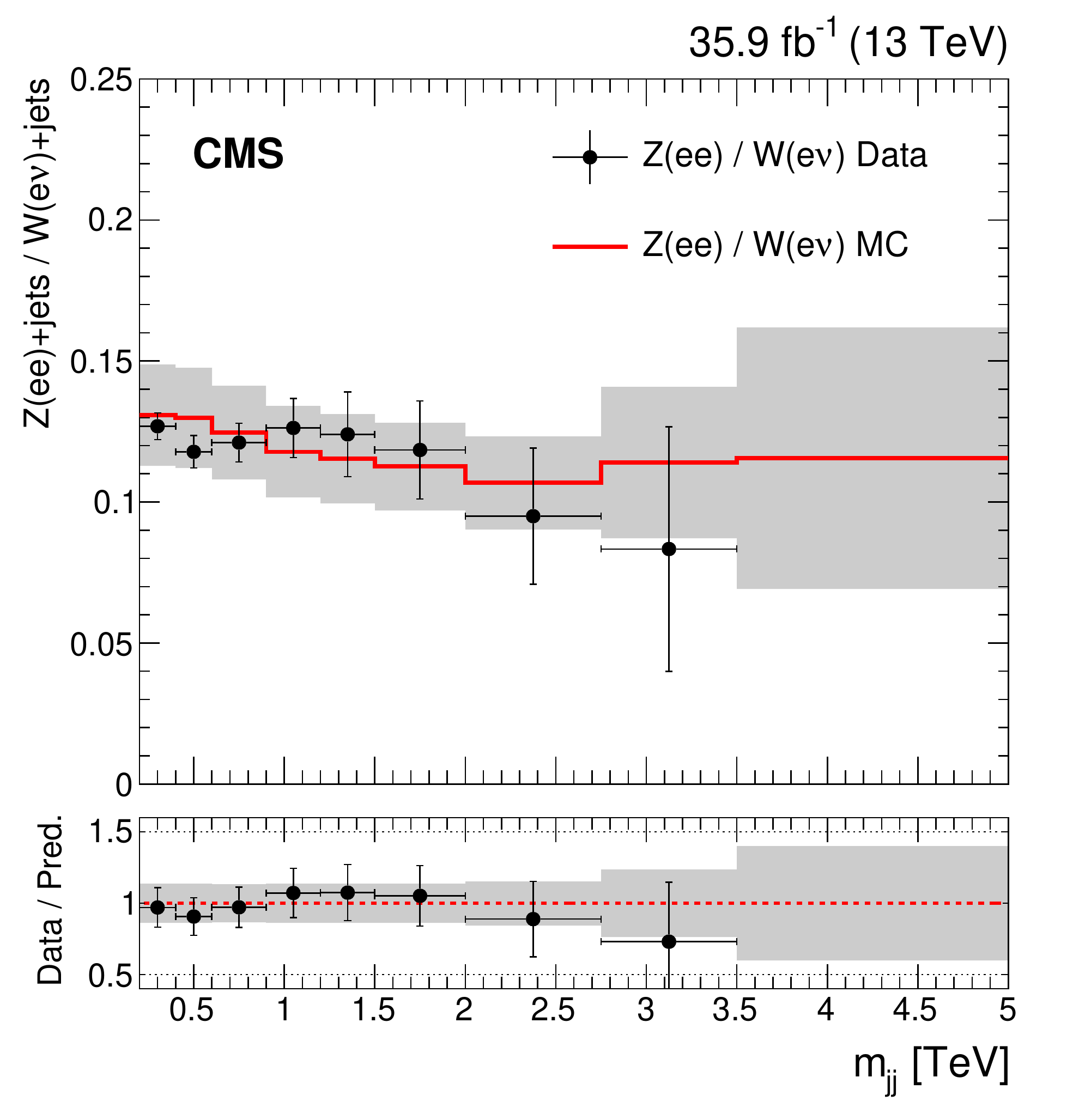}
\caption{Comparison between data and simulation of the \Zmmjets/\Wmnjets (left) and \Zeejets/\Wevjets (right) ratios as functions of \mjj, computed in the shape analysis phase-space. In the bottom panels, ratios of data with the pre-fit background prediction are reported. The gray bands include both the theoretical and experimental systematic uncertainties listed in Table~\ref{tab:vbfsysunc}, as well as the statistical uncertainty in the simulation.}
\label{fig:zw_ratio_validation}
\end{figure*}

To assess the level of agreement between data and simulation obtained through the application of \pt--\mjj dependent NLO corrections to both the \Vjets (QCD) and \Vjets (EW) processes, the ratio between the number of \Zjets and \Wjets events in the CRs in bins of \mjj is used as a benchmark. Figure~\ref{fig:zw_ratio_validation} shows the ratio of the \Zjets and \Wjets event yields in the muon (left) and electron (right) CRs, respectively. A good agreement is observed between data and simulation and local differences are covered by the systematic uncertainties listed in Table~\ref{tab:vbfsysunc}.

\subsection{Other backgrounds}

In addition to the \Vjets processes, several other minor sources of background contribute to the total event yield in the SR. These include QCD multijet events that typically have small genuine \ptmiss. However, jet momentum mismeasurements and instrumental effects may give rise to large \ptmiss tails. A $\text{min}\Delta\phi$ extrapolation method~\cite{Collaboration:2011ida} is used to estimate this background from data, where a QCD multijet enriched CR is defined by selecting events that fail the $\text{min}\Delta\phi$ requirement between the jets and the \ptvecmiss vector, but still fulfill the remaining event selection criteria.  A transfer factor, derived from simulated QCD multijet events, is used to estimate the background in the SR from the event rate measured in the low-$\dphijmet$ sample. The low-$\dphijmet$ region contains a significant contamination from \Vjets production, which have genuine \ptmiss. They contribute about 40\% of the total event yield for \mjj smaller than 500\GeV, and about 80\% for ${\mjj > 3\TeV}$. This contamination is estimated from simulation and subtracted from the event yield measured in the low-$\dphijmet$ sample. An uncertainty of 20\% is assigned while performing the subtraction, which results in an uncertainty of about 30\% in the estimated QCD multijet background in the SR. The MC statistical uncertainty of the QCD multijet samples, which affects the transfer factor prediction, is also considered and is found to vary between 40 and 100\% as a function of \mjj. Lastly, a validation of the $\Delta\phi$ method is performed using a purer sample of QCD multijet events that pass the analysis requirements, but have \ptmiss in the range of 100--175\GeV. In this validation region, the predicted QCD background is found to agree with the observation within 50\%, which is taken as a conservative estimate of an additional uncertainty.

The remaining background sources include top quark production and diboson processes, which are estimated from simulation. The \pt distribution of the top quark in simulation is corrected to match the observed \pt distribution in data~\cite{Czakon:2015owf}. An uncertainty of about 10\% is assigned to the overall top quark background normalization, while an additional 10\% uncertainty is added to account for the modeling of the top quark \pt distribution in simulation. The overall normalization of the diboson background has an uncertainty of about 15\%~\cite{Khachatryan:2016txa,Khachatryan:2016tgp}. The uncertainties in the top quark and diboson backgrounds are correlated across the SR and the CRs. Several experimental sources of uncertainty are also assigned to these backgrounds. An uncertainty of 2.5\% in the integrated luminosity measurement~\cite{CMS-PAS-LUM-17-001} is propagated to the background yields. The uncertainty in the efficiency of the \cPqb~quark jet veto is estimated to be around 3\% for the top quark background and of about 1\% for the other simulated processes. The uncertainty related to the jet energy scale varies between 8 and 15\%, depending on both the process and the CR.

\section{Results}\label{sec:results}

This section presents the results obtained from the shape and the cut-and-count analyses. These include 95\% \CL upper limits on \brhinv, and an interpretation of the search in the context of
a BSM model which allows for the presence of a SM-like Higgs boson with a mass between 110 and 1000\GeV.

\subsection{The shape analysis}\label{sec:shape_analysis_results}

\begin{figure*}[!htb]
\centering
\includegraphics[width=0.48\textwidth]{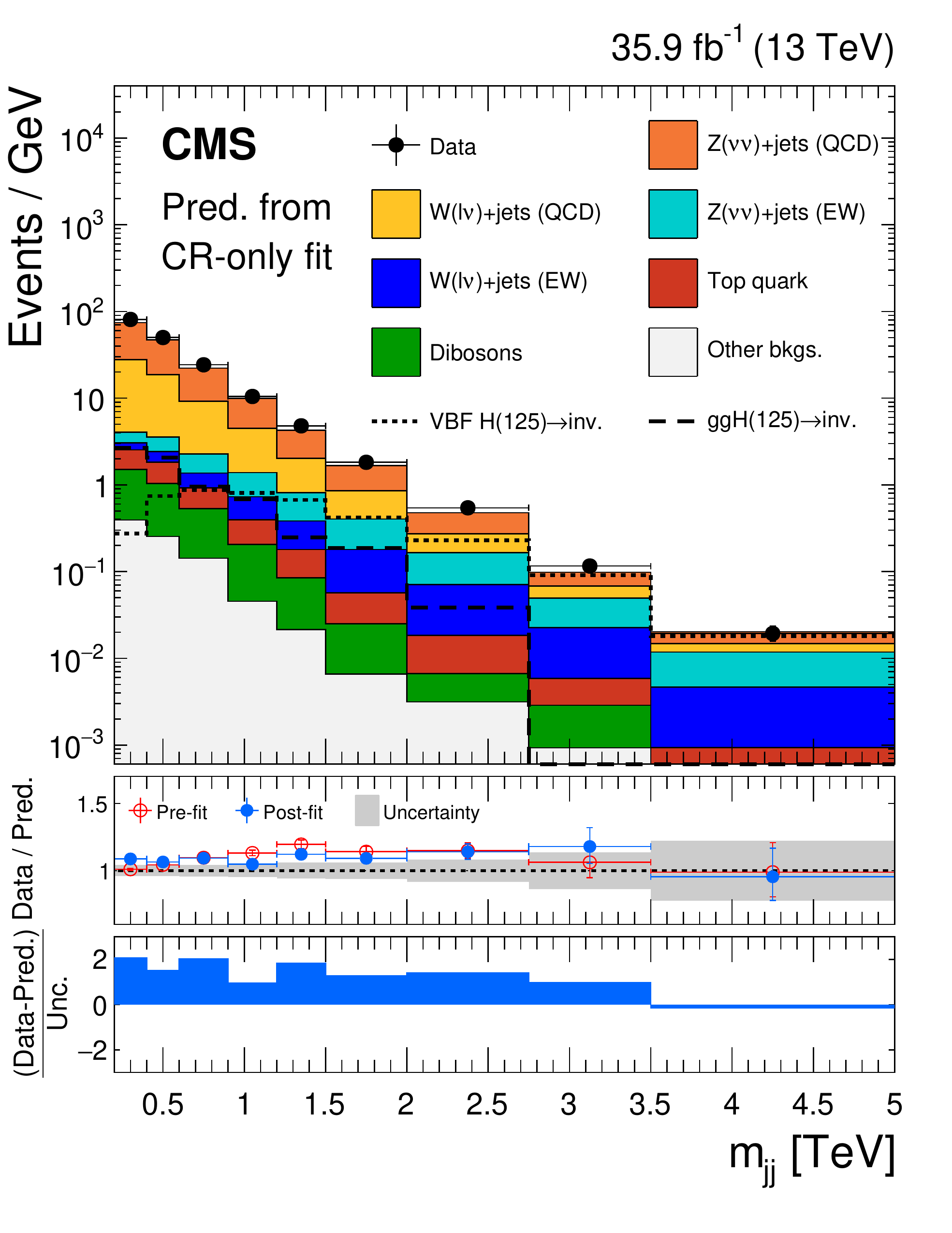}
\includegraphics[width=0.48\textwidth]{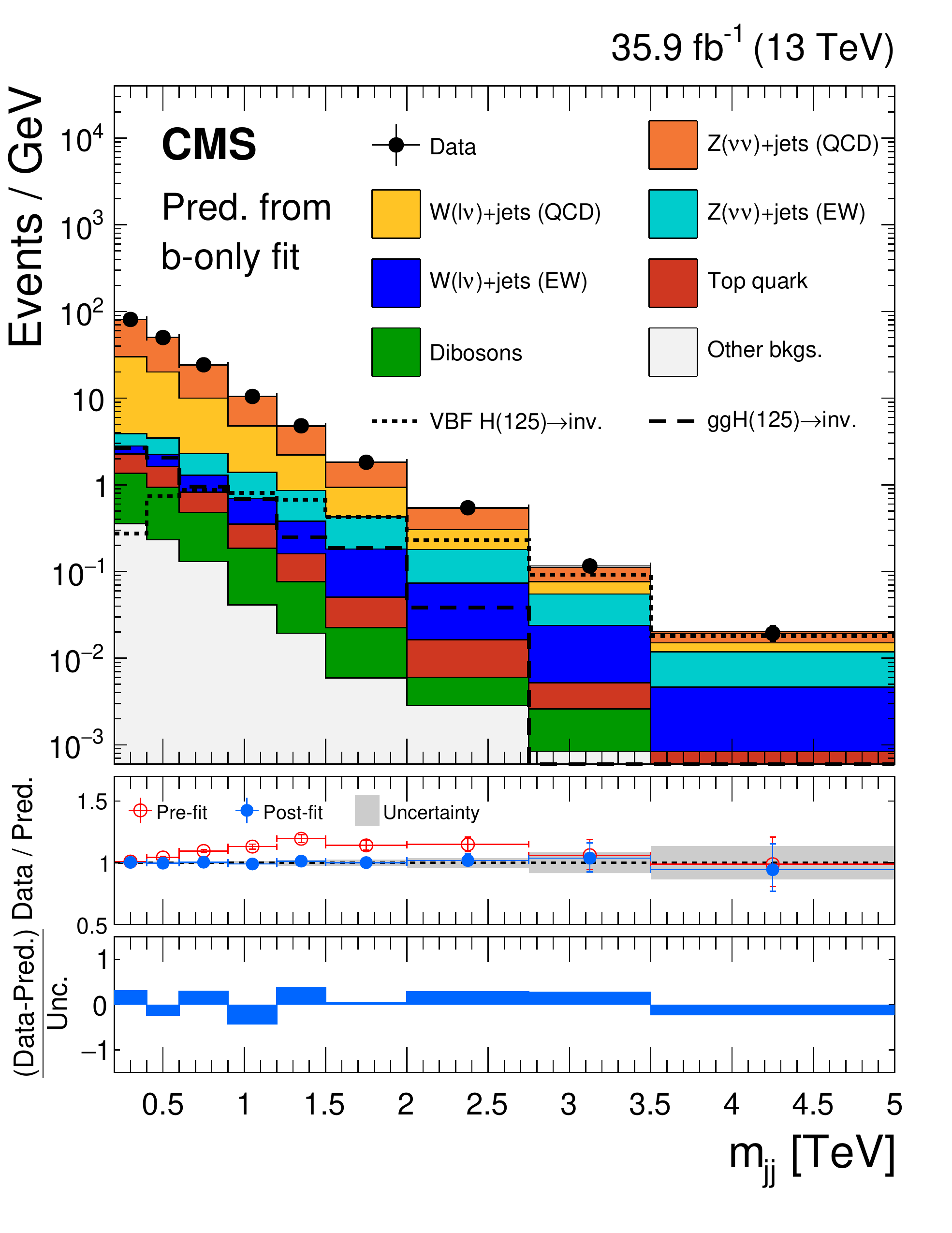}
\caption{The observed \mjj distribution of the shape analysis SR compared to the post-fit backgrounds from various SM processes. On the left, the predicted backgrounds are obtained from a combined fit to the data in all the CRs, but excluding the SR. On the right, the predicted backgrounds are obtained from a combined fit to the data in all the CRs, as well as in the SR, assuming the absence of any signal. Expected signal distributions for a 125\GeV Higgs boson produced through $\Pg\Pg\PH$ and VBF modes, and decaying to invisible particles with a branching fraction $\brhinv = 1$, are overlaid. The last bin includes all events with ${\mjj > 3.5\TeV}$. The description of the ratio panels is the same as in Fig.~\ref{fig:fit_controlregion}.}
\label{fig:mjj_signalregion}
\end{figure*}

The observed and the expected \mjj distributions in the SR, obtained after applying the full event selection, are shown in Fig.~\ref{fig:mjj_signalregion}. The background prediction shown in Fig.~\ref{fig:mjj_signalregion} (left) is obtained from a fit to the data in the CRs. Signal distributions for the SM Higgs boson produced via the $\Pg\Pg\PH$ and VBF modes are overlaid, assuming $\brhinv = 1$. The estimated background yields from the CR-only fit are listed in Table~\ref{tab:mjj_signalregion}, along with the observed event yield in the SR. The large contamination from $\Pg\Pg\PH$ production arises from the low-\mjj bins, which represent the least sensitive region to \hinv decays. Systematic uncertainties in the \Vjets transfer factors and in the minor backgrounds introduce correlations across the \mjj bins used in the fit. The correlations in the predicted background yields in each \mjj bin are reported in Table~\ref{tab:correlation}.

\begin{table*}[!htb]
\centering
\topcaption{Expected event yields in each \mjj bin for various background processes in the SR of the shape analysis. The background yields and the corresponding uncertainties are obtained after performing a combined fit across all the CRs, but excluding data in the SR. The ``other backgrounds'' includes QCD multijet and \Zlljets processes. The expected total signal contribution for the 125\GeV Higgs boson, decaying to invisible particles with a branching fraction $\brhinv = 1$, and the observed event yields are also reported.}
\resizebox{\textwidth}{!}{
\begin{tabular}{l c c c c c c c c c}
\hline
 \multicolumn{1}{l}{Process} & \multicolumn{8}{c}{\mjj range in \TeV} \\ [\cmsTabSkip]
                     & 0.2--0.4 & 0.4--0.6 & 0.6--0.9 & 0.9--1.2 & 1.2--1.5 & 1.5--2.0 & 2.0--2.75 & 2.75--3.5 & $>$ 3.5 \\
\hline
$\PZ(\nu\nu)$ (QCD)  & 9311 $\pm$ 388  & 5669  $\pm$ 257  & 3884 $\pm$ 179  & 1648 $\pm$ 88  & 677  $\pm$ 42  & 405  $\pm$ 28  & 153  $\pm$ 14  & 22.8 $\pm$ 3.5 & 8.1  $\pm$ 2.2 \\
$\PZ(\nu\nu)$ (EW)   & 201  $\pm$ 8    & 228   $\pm$ 10   & 273  $\pm$ 13   & 198  $\pm$ 11  & 129  $\pm$ 8   & 112  $\pm$ 8   & 70.6 $\pm$ 6.6 & 20.2 $\pm$ 3.1 & 10.8 $\pm$ 2.9 \\
$\PW(\ell\nu)$ (QCD) & 4755 $\pm$ 267  & 3017  $\pm$ 180  & 2090 $\pm$ 130  & 928  $\pm$ 63  & 361  $\pm$ 28  & 227  $\pm$ 19  & 80.4 $\pm$ 9.1 & 13.7 $\pm$ 2.7 & 4.5  $\pm$ 1.9 \\
$\PW(\ell\nu)$ (EW)  & 102  $\pm$ 14   & 118   $\pm$ 16   & 133  $\pm$ 18   & 100  $\pm$ 13  & 61.2 $\pm$ 8.1 & 61.4 $\pm$ 7.6 & 39.4 $\pm$ 4.9 & 12.6 $\pm$ 1.9 & 5.6 $\pm$ 1.4 \\
Top quark            & 208  $\pm$ 37   & 159   $\pm$ 28   & 119  $\pm$ 21   & 57.6 $\pm$ 10.2 & 28.7 $\pm$ 5.1 & 16.1 $\pm$ 2.9 & 8.9 $\pm$ 1.6 & 2.2  $\pm$ 0.4 & 0.7 $\pm$ 0.1 \\
Dibosons             & 222  $\pm$ 39   & 157   $\pm$ 28   & 116  $\pm$ 21   & 48.2 $\pm$ 8.5 & 19.0 $\pm$ 3.4 & 9.3 $\pm$ 1.6  & 2.6 $\pm$ 0.5  & 1.4 $\pm$ 0.3  & 0.4 $\pm$ 0.1 \\
Others               & 78.6 $\pm$ 19.5 & 51.0 $\pm$ 11.6  & 42.8 $\pm$ 11.5 & 13.6 $\pm$ 2.9 & 6.5 $\pm$ 1.5  & 3.3 $\pm$ 0.8  & 2.4 $\pm$ 0.6  & 0.7 $\pm$ 0.2  & 0.3 $\pm$ 0.4 \\ [\cmsTabSkip]
Total bkg.           & 14878 $\pm$ 566 & 9401 $\pm$ 387   & 6658 $\pm$ 271  & 2994 $\pm$ 144 & 1283 $\pm$ 69  & 834 $\pm$ 51   & 358 $\pm$ 29 & 73.8 $\pm$ 9.4   & 30.3 $\pm$ 7.4 \\ [\cmsTabSkip]
Signal               & 590 $\pm$ 244   & 559 $\pm$ 199    & 547 $\pm$ 151   & 447 $\pm$ 109  & 276 $\pm$ 58   & 304 $\pm$ 66   & 201 $\pm$ 36 & 68.6 $\pm$ 11.7  & 30.0 $\pm$ 6.4 \\ [\cmsTabSkip]
Data                 & 16177           & 10008            & 7277            & 3138           & 1439           & 911            & 408          & 87               & 29  \\
\hline
\end{tabular}
}
\label{tab:mjj_signalregion}
\end{table*}

An excess of about 4--10\% is observed in the SR data when compared to the estimated backgrounds. The discrepancy resides mainly in the bulk of the \mjj distribution. The shape of the excess is inconsistent with the characteristic features of a VBF signal, whose presence is expected to produce an increasing discrepancy between data and backgrounds as \mjj increases. A goodness of fit test, based on a saturated $\chi^{2}$ test statistic~\cite{Baker:1983tu,lindsey1996parametric}, yields a \textit{p}-value of about 6\% indicating that the data are compatible with the SM prediction.

Figure~\ref{fig:mjj_signalregion} (right) shows the background prediction obtained after including events from the SR in the fit, but assuming the absence of a signal. Such a fit is referred to as the ``b-only fit''. The comparison between the results of the b-only fit with that allowing for the presence of the signal is used to set an upper limit on \brhinv. In the b-only fit, the \Vjets estimate in the SR can vary with respect to the prediction from the CRs within the systematic uncertainties assigned to the transfer factors. Therefore, the additional constraint due to the data in the SR mitigates the excess shown in Fig.~\ref{fig:mjj_signalregion} (left), yielding a \textit{p}-value for the b-only fit of about 65\%.

The results of this search are interpreted in terms of an upper limit on the product of the Higgs boson production cross section and its branching fraction to invisible particles, $\sigma\brhinv$, relative to the predicted cross section assuming SM interactions, $\sigma_{\text{SM}}$. Observed and expected 95\% \CL upper limits are computed using an asymptotic approximation of the \CLs  method~\cite{Junk:1999kv,Read:2002av} with a profile likelihood ratio test statistic~\cite{Cowan:2010js} in which systematic uncertainties are modeled as nuisance parameters following a frequentist approach~\cite{CMS-NOTE-2011-005}. The profile likelihood ratio is defined as:
\begin{linenomath}
\ifthenelse{\boolean{cms@external}}{
\begin{equation}
\begin{aligned}
q  &= -2 \Delta \ln \mathcal{L} \\
   &= -2 \ln \frac{\mathcal{L}(\text{data}|\sigmabr,\boldsymbol{\hat{\theta}_{a}},\boldsymbol{\hat{\kappa}_{a}})}{\mathcal{L}(\text{data}|(\sigma/\sigma_{\mathrm{SM}}) \, \hat{\mathcal{B}}(\PH\to \text{inv}),\boldsymbol{\hat{\theta}},\boldsymbol{\hat{\kappa}})}
\end{aligned}
\end{equation}
}{
\begin{equation}
q  = -2 \Delta \ln \mathcal{L} = -2 \ln \frac{\mathcal{L}(\text{data}|\sigmabr,\boldsymbol{\hat{\theta}_{a}},\boldsymbol{\hat{\kappa}_{a}})}{\mathcal{L}(\text{data}|(\sigma/\sigma_{\mathrm{SM}}) \, \hat{\mathcal{B}}(\PH\to \text{inv}),\boldsymbol{\hat{\theta}},\boldsymbol{\hat{\kappa}})}
\end{equation}
}
\end{linenomath}
where ${(\sigma/\sigma_{\mathrm{SM}})\,\hat{\mathcal{B}}(\PH\to\text{inv})}$ represents the value of the signal strength that maximizes the likelihood $\mathcal{L}$ for the data, while $\boldsymbol{\hat{\theta}}~(\boldsymbol{\hat{\kappa}})$ and $\boldsymbol{\hat{\theta}_{a}}~(\boldsymbol{\hat{\kappa}_{a}})$ denote the best fit estimates for the nuisance parameters (\Zvvjets rate in each bin) and the estimates for a given fixed value of ${\sigmabr}$, respectively.

The relative contributions of the VBF and $\Pg\Pg\PH$ production modes are fixed to the SM prediction within their uncertainties. The uncertainties in the predictions of the inclusive VBF and $\Pg\Pg\PH$ production cross sections due to PDF uncertainties, renormalization and factorization scale variations are taken from Ref.~\cite{deFlorian:2016spz}. An additional uncertainty of 40\% is assigned to the expected $\Pg\Pg\PH$ contribution. This accounts for both the limited knowledge of the $\Pg\Pg\PH$ cross section in association with two or more jets, as well as the uncertainty in the prediction of the $\Pg\Pg\PH$ differential cross section for large Higgs boson transverse momentum, $\pt^{\PH} > 250 \GeV$.  The former contribution is obtained by following the recipe outlined in Ref.~\cite{deFlorian:2016spz} and is found to be about 30\%, while the latter uncertainty is estimated by comparing the prediction from \textsc{powheg+\MINLO}~\cite{Hamilton:2012np} with the one from \MGvATNLO and ranges between 20 and 25\%. Furthermore, the uncertainties in the signal acceptance due to the choice of the PDF set are also evaluated independently for the different signal processes, and are treated as independent nuisance parameters in the fit.

The observed (expected) 95\% \CL upper limit on \brhinv is measured to be 0.33 (0.25), and the regions containing 68\% and 95\% of the distribution of upper limits, expected in absence of a signal, are found to be 0.18--0.35 and 0.14--0.47, respectively.

The background estimates reported in Table~\ref{tab:mjj_signalregion}, along with the correlation matrix presented in Table~\ref{tab:correlation}, can be used in the simplified likelihood approach detailed in Ref.~\cite{CMS-NOTE-2017-001} to reinterpret these results in theoretical models different from those presented in this Letter.

\begin{table*}[!htb]
\centering
\topcaption{Correlation between the uncertainties in predicted background yields across the \mjj bins of the shape analysis SR. The backgrounds are estimated by fitting the data in the CRs. Bin ranges are expressed in\TeV.}
\cmsTable{
\begin{tabular}{l c c c c c c c c c}
\hline
\multicolumn{10}{c}{Correlation coefficients}\\[\cmsTabSkip]
& 0.2--0.4 & 0.4--0.6 & 0.6--0.9 & 0.9--1.2 & 1.2--1.5 & 1.5--2.0 & 2.0--2.75 & 2.75--3.5 & $>3.5$ \\
\hline
 0.2--0.4  & 1.00 & \NA  & \NA  & \NA  & \NA  & \NA  & \NA  & \NA  & \NA \\
 0.4--0.6  & 0.88 & 1.00 & \NA  & \NA  & \NA  & \NA  & \NA  & \NA  & \NA \\
 0.6--0.9  & 0.85 & 0.84 & 1.00 & \NA  & \NA  & \NA  & \NA  & \NA  & \NA \\
 0.9--1.2  & 0.78 & 0.76 & 0.75 & 1.00 & \NA  & \NA  & \NA  & \NA  & \NA \\
 1.2--1.5  & 0.70 & 0.72 & 0.71 & 0.60 & 1.00 & \NA  & \NA  & \NA  & \NA \\
 1.5--2.0  & 0.62 & 0.57 & 0.63 & 0.59 & 0.54 & 1.00 & \NA  & \NA  & \NA \\
 2.0--2.75 & 0.38 & 0.40 & 0.43 & 0.43 & 0.45 & 0.43 & 1.00 & \NA  & \NA \\
 2.75--3.5 & 0.28 & 0.28 & 0.33 & 0.26 & 0.34 & 0.27 & 0.22 & 1.00 & \NA \\
 $>$3.5   & 0.22 & 0.21 & 0.22 & 0.23 & 0.24 & 0.20 & 0.23 & 0.19 & 1.00 \\
\hline
\end{tabular}
}
\label{tab:correlation}
\end{table*}

\subsection{The cut-and-count analysis}\label{sec:cut_and_count_results}

The cut-and-count analysis is presented because it allows for an easier reinterpretation of this search in the context of other theoretical models that predict \ptmiss plus VBF dijet signatures. The observed event yield after the cut-and-count selection is reported in Table~\ref{tab:cc_signalregion}, along with the predicted backgrounds in the SR. The backgrounds are estimated by fitting the data in the CRs. An excess, characterized by a significance of about 2.5 standard deviations, is observed in the SR compared to the background prediction obtained from the CRs. As for the shape analysis, this excess is mostly due to low \mjj events. The excess is incompatible with a VBF Higgs boson signal and, upon detailed scrutiny, is ascribed to a statistical fluctuation.

The results of the cut-and-count analysis are presented in terms of a 95\% CL upper limit on \brhinv using the statistical procedure outlined in Section~\ref{sec:shape_analysis_results}. The observed (expected) upper limit is found to be 0.58 (0.30), and the regions containing 68\% and 95\% of the distribution of upper limits, expected in absence of a signal, are found to be 0.22--0.43 and 0.17--0.58, respectively.

\begin{table*}[!htb]
\centering
\topcaption{Expected event yields in the SR and in the CRs of the cut-and-count analysis for various SM processes. The background yields and the corresponding uncertainties are obtained from a combined fit to data in all the CRs, but excluding data in the SR. The expected total signal contribution for the 125\GeV Higgs boson, decaying to invisible particles with a branching fraction $\brhinv = 1$, and the observed event yields are also reported.}
\cmsTable{
\begin{tabular}{l c c c c c}
  \hline
  Process             & Signal region & Dimuon CR & Dielectron CR & Single-muon CR & Single-electron CR \\
  \hline
  $\PZ(\nu\nu)$ (QCD)   & $810 \pm 71$   & \NA              & \NA              & \NA               & \NA \\
  $\PZ(\nu\nu)$ (EW)    & $269 \pm 33$   & \NA              & \NA              & \NA               & \NA \\
  $\PZ(\ell\ell)$ (QCD) & \NA            & $91.5 \pm 7.6$ & $66.5 \pm 6.0$ & $27.1 \pm 1.2$  & $5.2 \pm 0.3$ \\
  $\PZ(\ell\ell)$ (EW)  & \NA            & $32.5 \pm 4.1$ & $24.1 \pm 3.2$ & $5.7 \pm  0.3$  & $2.4 \pm 0.2$ \\
  $\PW(\ell\nu)$ (QCD)  & $499 \pm 33$   & $0.2 \pm 0.2$    & $0.9 \pm 0.6$  & $907 \pm 30$    & $544 \pm 21$ \\
  $\PW(\ell\nu)$ (EW)   & $141 \pm 11$   & $0.1 \pm 0.1$    & \NA            & $406 \pm 15$    & $254 \pm 11$ \\
  Top quark           & $37.8 \pm 8.8$   & $4.8 \pm 1.4$    & $2.9 \pm 1.0$  & $112 \pm 22$        & $74.2 \pm 13.6$ \\
  Dibosons            & $18.6 \pm 6.2$   & $2.3 \pm 1.1$    & $0.7 \pm 0.4$  & $21.3 \pm 4.4$      & $14.4 \pm 3.7$ \\
  Others              & $3.3 \pm 2.3$    & \NA              & \NA             & $22.9 \pm 13.9$    & $2.1 \pm 1.9$ \\ [\cmsTabSkip]
  Total bkg.          & $1779 \pm 96$    & $131 \pm 8$      & $95.0 \pm 6.3$ & $1502 \pm 34$       & $896 \pm 24$ \\ [\cmsTabSkip]
  Signal $m_{\PH} = 125.09 \GeV$ & $743 \pm 129 $   & \NA & \NA & \NA & \NA \\ [\cmsTabSkip]
  Data                & 2035           & 114            & 104            & 1504            & 902 \\
  \hline
\end{tabular}
}
\label{tab:cc_signalregion}
\end{table*}

\subsection{Constraints on a SM-like Higgs boson}

\begin{figure*}[htb]
\centering
\includegraphics[width=0.45\textwidth]{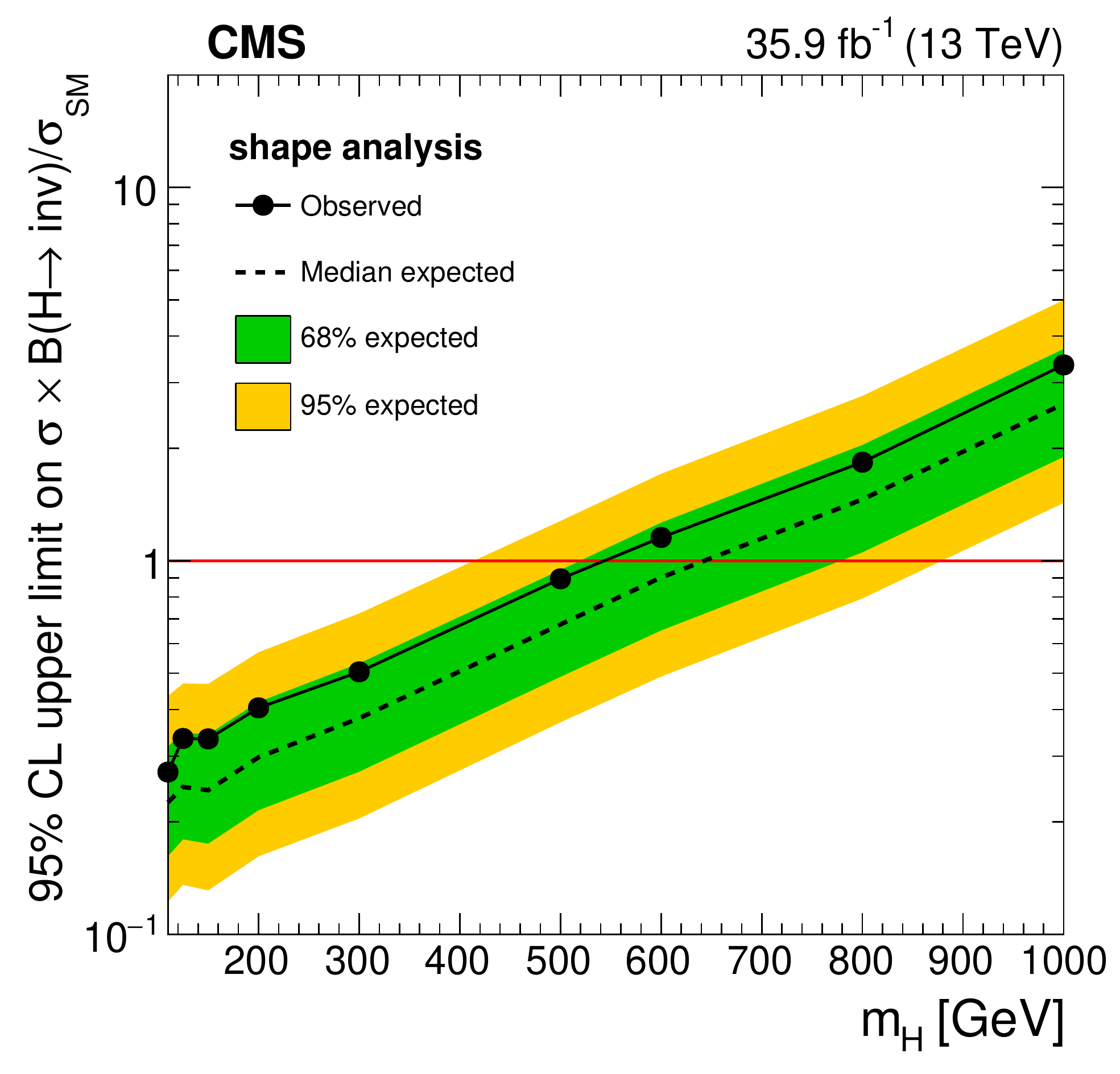}
\includegraphics[width=0.45\textwidth]{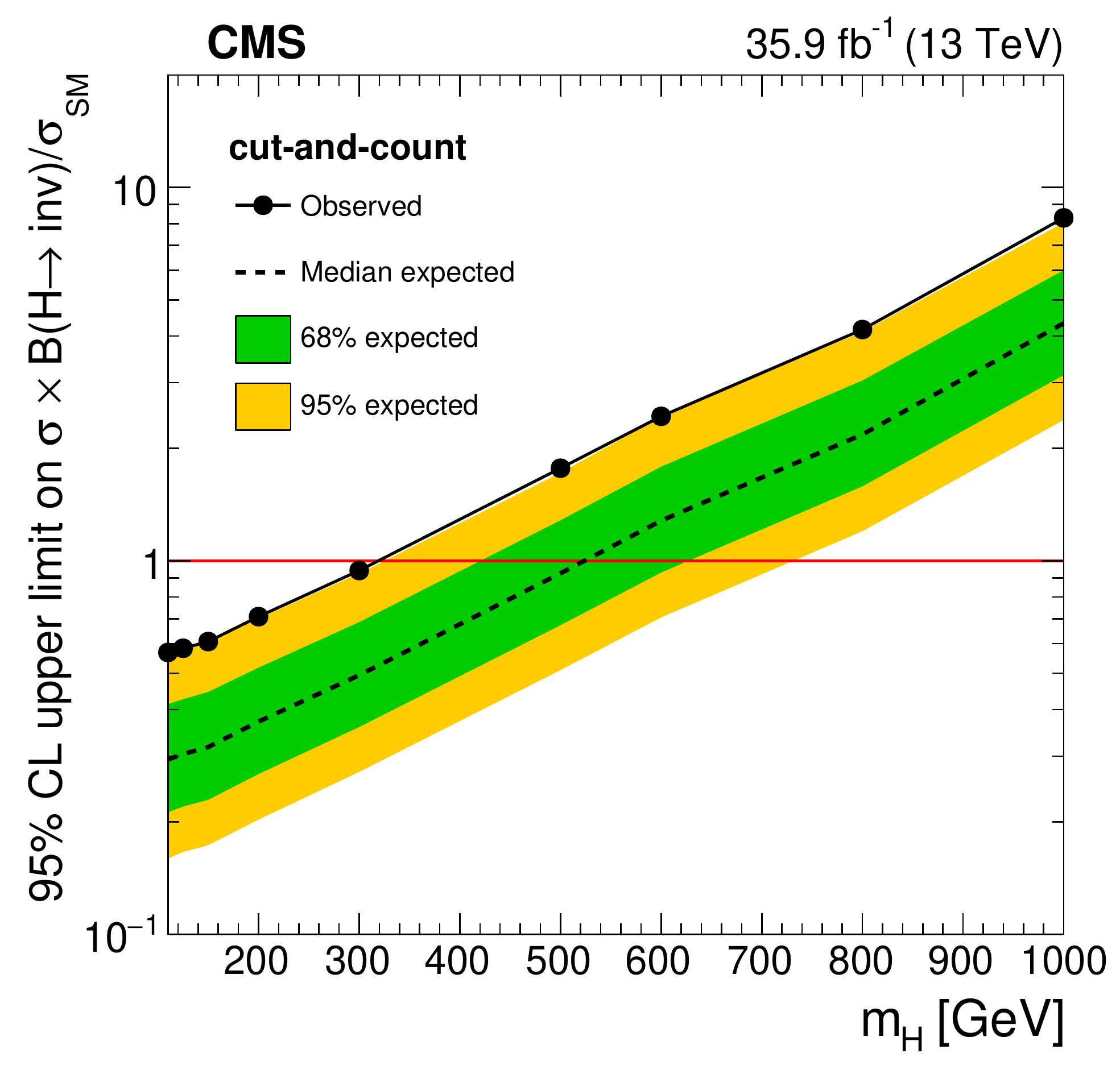}
\caption{Expected and observed 95\% \CL upper limits on ${\sigmabr}$ for an SM-like Higgs boson as a function of its mass ($m_{\PH}$). On the left, observed (solid black) and expected (dashed black) upper limits are obtained from the shape analysis while, on the right, results from the cut-and-count analysis are reported. The 68\% (green) and 95\% (yellow) CL intervals around the expected upper limits are also shown for both the shape and the cut-and-count analyses.}
\label{fig:xsec_limits}
\end{figure*}

The results presented are also interpreted in the context of an additional SM-like Higgs boson that does not mix with the 125\GeV boson and decays to invisible particles~\cite{Heinemeyer:2013tqa}. Such a boson may be produced via both the $\Pg\Pg\PH$ and VBF mechanisms. This model has already been studied in earlier CMS publications~\cite{Chatrchyan:2014tja,Khachatryan:2015cwa,Sirunyan:2017qfc}. Upper limits, computed at 95\% \CL on ${\sigmabr}$, are shown in Fig.~\ref{fig:xsec_limits} as a function of the SM-like Higgs boson mass hypothesis ($m_{\PH}$) for both the shape and the cut-and-count analyses.

\section{Combined limits on \texorpdfstring{\hinv}{H to invisible} from 2016 data}\label{sec:combination_2016}

The common feature of all the searches included in this combination is a large \ptmiss, where at least one high-\pt jet or a weak boson recoils against the invisible particles produced by the Higgs boson decay. Specific topological selections are designed to reduce the contamination from large SM backgrounds, targeting a particular Higgs boson production mode. The analyses included in this combination are listed in Table~\ref{tab:analysis_combination}, together with their expected signal composition and their individual upper limits on \brhinv. The results quoted for the VBF channel come from the shape analysis described earlier in this Letter. The $\PZ(\ell^{+}\ell^{-})\PH$ analysis is identical to the one described in Ref.~\cite{Sirunyan:2017qfc}, where the expected signal comes entirely from invisible decays of the SM Higgs boson produced in association with a leptonically decaying \PZ boson, via either ${\cPq\cPq \to \PZ\PH}$ or ${\Pg\Pg \to \PZ\PH}$ production. In contrast, the $\PV(\cPq\cPq\textsf{'})\PH$ and the $\Pg\Pg\PH$-tagged searches are similar to those described in Ref.~\cite{Sirunyan:2017jix}, but events which overlap with the VBF analysis have been removed to avoid double counting. In both the $\Pg\Pg\PH$ and $\PV(\cPq\cPq\textsf{'})\PH$ searches, overlapping events represent about 6\,(15)\% of the total background for a \ptmiss of about 250\,(1000)\GeV. The overlap removal introduces a 5\% loss in the expected exclusion sensitivity compared to that of Ref.~\cite{Sirunyan:2017jix}. Both the $\PV(\cPq\cPq\textsf{'})\PH$ and the $\Pg\Pg\PH$ searches target events with at least one high-\pt central jet, and their SRs contain a mixture of different production modes. This mixture results from the limited discrimination power of the substructure observables exploited to select boosted $\PV(\cPq\cPq\textsf{'})\PH$ candidates.

\begin{table*}[htb]
\centering
\topcaption{Signal composition and upper limits (observed and expected) on the invisible Higgs boson branching fraction classified according to the final state considered in each analysis. The relative contributions from the different Higgs production mechanisms are derived from simulation, fixing the Higgs boson mass to 125.09\GeV and assuming SM production cross sections.}
\cmsTable{
\begin{tabular}{l l c c c}
\hline
Analysis       & Final state  & Signal composition  & Observed limit & Expected limit\\
\hline
VBF-tag                    & $\text{VBF-jet}+\ptmiss$                & 52\% VBF, 48\% $\Pg\Pg\PH$   & 0.33 & 0.25 \\ [\cmsTabSkip]
\multirow{2}{*}{$\PV\PH$-tag}  & $\Zll+\ptmiss$~\cite{Sirunyan:2017qfc}    & 79\% $\cPq\cPq\PZ\PH$, 21\% $\Pg\Pg\PZ\PH$ & 0.40 & 0.42 \\
                           & $\PV(\cPq\cPq\textsf{'})+\ptmiss$~\cite{Sirunyan:2017jix} & 39\% $\Pg\Pg\PH$, 6\% VBF, 33\% $\PW\PH$, 22\% $\PZ\PH$ & 0.50 & 0.48\\ [\cmsTabSkip]
$\Pg\Pg\PH$-tag            & $\text{jets}+\ptmiss$~\cite{Sirunyan:2017jix}     & 80\% $\Pg\Pg\PH$, 12\% VBF, 5\% $\PW\PH$, 3\% $\PZ\PH$ & 0.66 & 0.59\\
\hline
\end{tabular}
}
\label{tab:analysis_combination}
\end{table*}

No significant deviations from the SM expectations are observed in any of the searches. The results are interpreted as an upper limit on ${\sigmabr}$. These limits are calculated following the same approach described in Section~\ref{sec:shape_analysis_results}. The combined likelihood fit accounts for correlations between the nuisance parameters in each search. The uncertainties in the diboson backgrounds (except for those considered in the ${\PZ(\ell\ell)\PH}$ channel), \ttbar and single top quark cross sections, lepton efficiencies, momentum scales, integrated luminosity, \cPqb~quark jet and \tauh vetoes are correlated among all the searches. In addition, the uncertainties in the inclusive signal production cross sections, due to renormalization and factorization scale variations, and the PDF uncertainties are also correlated across the channels. In contrast, since the jet kinematics in the VBF search differ from that in the other analyses, jet energy scale and resolution uncertainties are correlated only across the $\Pg\Pg\PH$ and $\PV\PH$-tagged categories. The theoretical uncertainties applied to the \Vjets (QCD) ratios are assumed to be uncorrelated between the VBF analysis and the other searches.

\begin{figure*}[htb]
\centering
\includegraphics[width=0.45\textwidth]{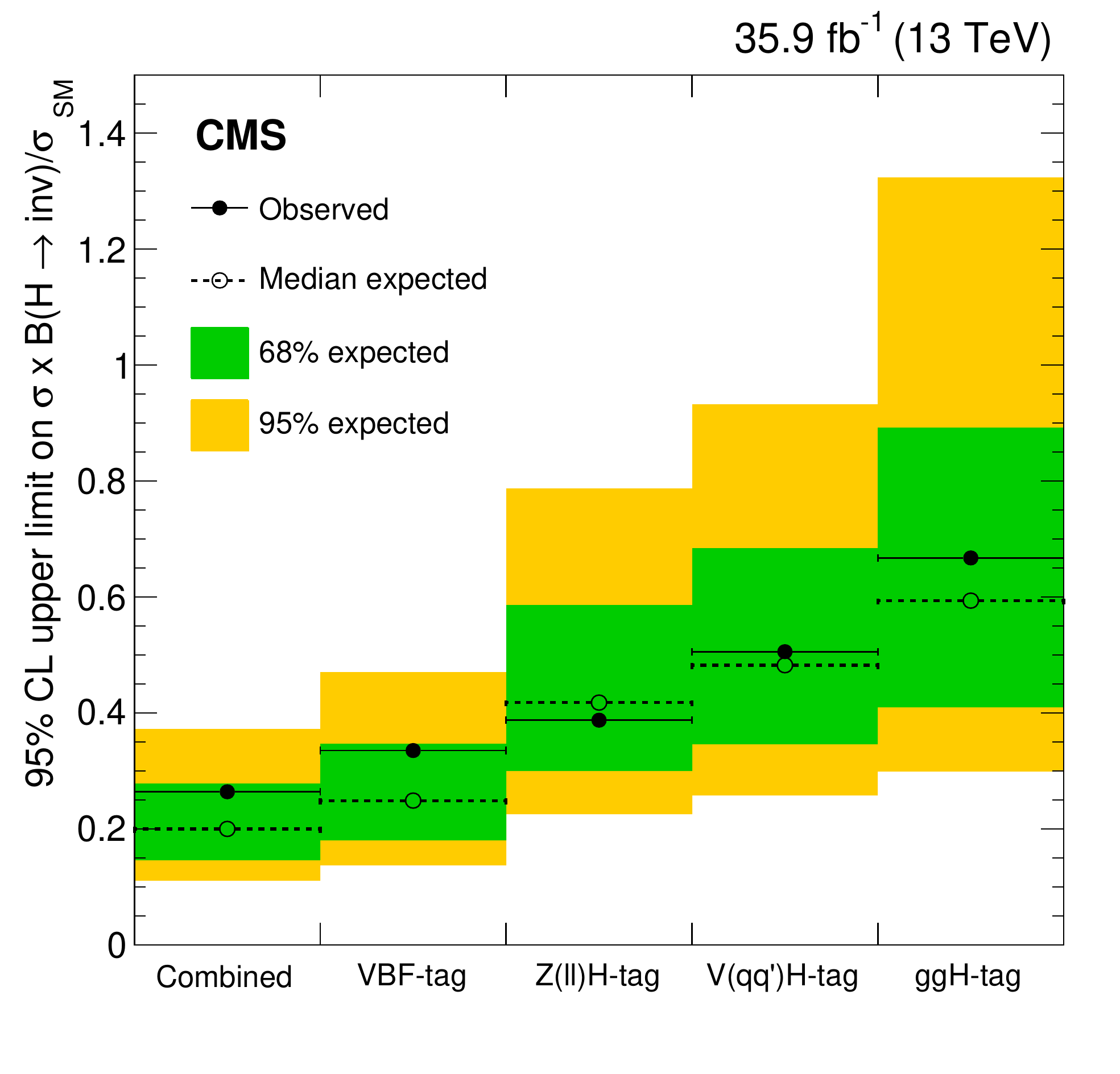}
\includegraphics[width=0.45\textwidth]{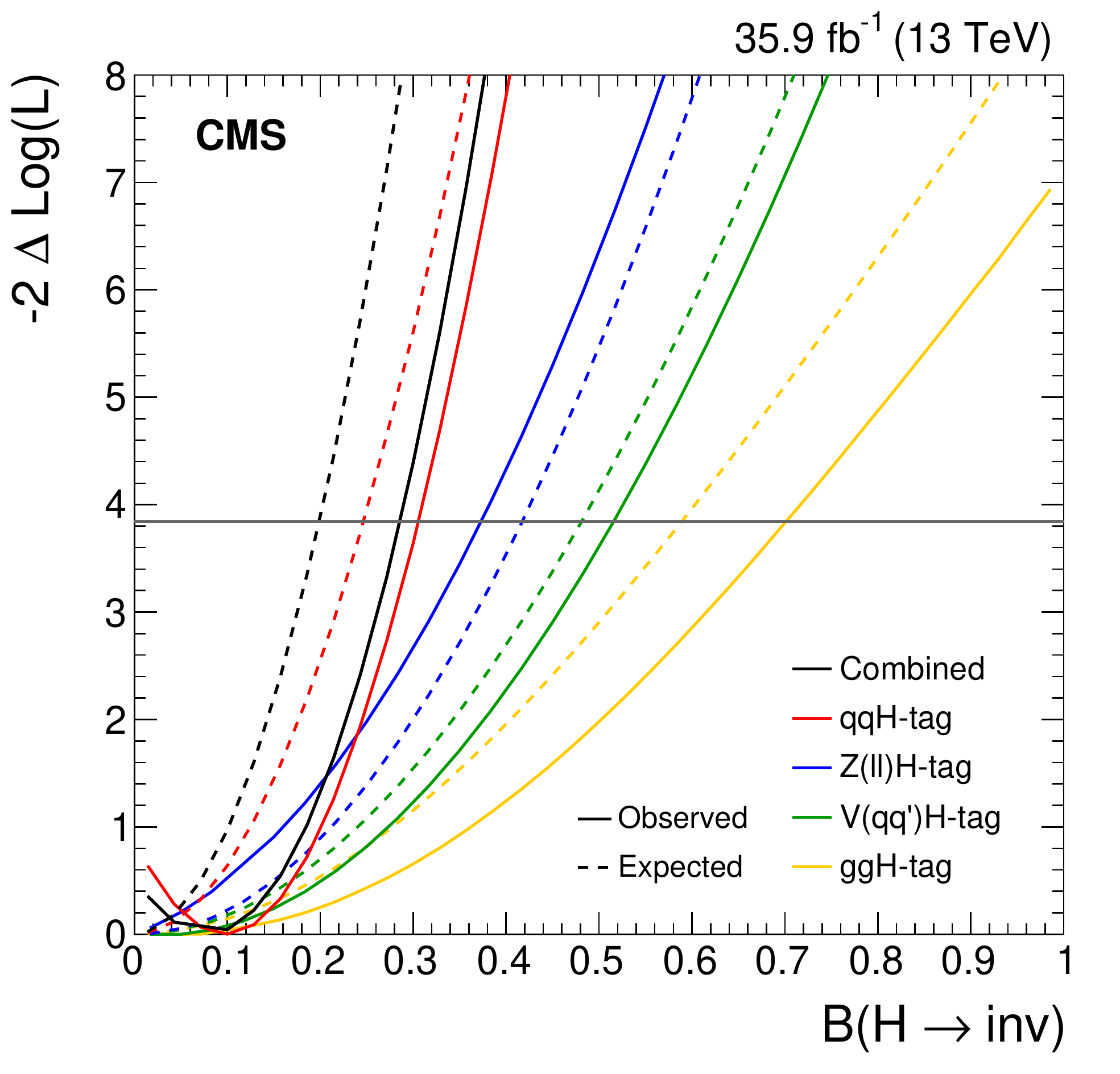}
\caption{On the left, observed and expected 95\% \CL upper limits on ${\sigmabr}$ for both individual categories targeting VBF, ${\PZ(\ell\ell)\PH}$, $\PV(\cPq\cPq\textsf{'})$\PH, and $\Pg\Pg\PH$ production mode, as well as their combination, assuming an SM Higgs boson with a mass of 125.09\GeV. On the right, profile likelihood ratios as a function of \brhinv. The solid curves represent the observations in data, while the dashed lines represent the expected result from a b-only fit. The observed and expected likelihood scans are reported for the full combination, as well as for the individual VBF, $\PZ(\ell\ell)\PH$, $\PV(\cPq\cPq')\PH$ and $\Pg\Pg\PH$-tagged analyses.}
\label{fit:combination_limit}
\end{figure*}

Observed and expected upper limits on ${\sigmabr}$ are computed at 95\% \CL and are presented in Fig.~\ref{fit:combination_limit} (left). Assuming SM cross sections for each production mode, the combination yields an observed (expected) upper limit of ${\brhinv < 0.26\,(0.20)}$. The profile likelihood ratios as a function of \brhinv, for both the combined fit and each individual search channel, are reported in Fig.~\ref{fit:combination_limit} (right). Results are shown for both data and an Asimov dataset~\cite{Cowan:2010js}, defined by fixing the nuisances parameters to their maximum likelihood estimate obtained from a fit to the data in which ${\brhinv = 0}$ is assumed.

\section{\texorpdfstring{Combination of 7, 8, and 13\TeV searches for \hinv decays}{Combination of 7, 8, and 13 TeV searches for H to invisible decays}}\label{sec:combination}

The analyses previously described and listed in Table~\ref{tab:analysis_combination} are further combined with earlier searches performed using data collected at $\sqrt{s}=7$, 8, and 13\TeV up to the end of 2015, as reported in Refs.~\cite{Khachatryan:2016whc,Chatrchyan:2014tja,Khachatryan:2016mdm}. The 7 and 8\TeV data, collected in 2011 and 2012, correspond to integrated luminosities of up to 4.9 and 19.7\fbinv~\cite{CMS-PAS-SMP-12-008,CMS-PAS-LUM-13-001}, respectively.  The 13\TeV data collected in 2015 correspond to an integrated luminosity of 2.3\fbinv~\cite{CMS-PAS-LUM-15-001}. Systematic uncertainties in the inclusive $\Pg\Pg\PH$, VBF, and $\PV\PH$ production cross sections are fully correlated across the 7, 8, and 13\TeV analyses. The uncertainty in the prediction of the Higgs boson \pt distribution in $\Pg\Pg\PH$ production, included in both the $\Pg\Pg\PH$ and $\PV\PH$ channels, and those arising from the limited knowledge of the PDFs, are also correlated among all searches. The uncertainties in the lepton and photon reconstruction and identification efficiencies, in the lepton momentum scales, and in the veto efficiency of leptons, \tauh candidates, and b quark jets are uncorrelated between 7+8 and 13\TeV searches. Similarly, uncertainties in the jet energy scale and resolution, mistag rate of leptons, and modeling of the unclustered particles are also uncorrelated between 7+8 and 13\TeV searches. The b jet energy scale and resolution uncertainties for the $\PZ(\cPqb\cPqb)\PH$ analysis are estimated using different techniques, and therefore are treated as uncorrelated with other channels. Theoretical uncertainties affecting the ratio of \Zvv and \Wlv predictions in the VBF searches, for both QCD and EW \Vjets processes, are uncorrelated across data sets because different strategies are followed to quantify and assign these uncertainties. In contrast, those affecting the \Zvv/\Wlv and \Zvv/\phojets ratios in the $\Pg\Pg\PH$ and $\PV(\cPq\cPq)\PH$ channels are correlated across the 7+8 and the 13\TeV (2015 data) searches, as described in Ref.~\cite{Khachatryan:2016whc}. The uncertainties in the tune of the underlying event simulation and in the pileup modeling are uncorrelated between 7+8 and 13\TeV searches. Finally, theoretical uncertainties affecting diboson and top quark cross sections, except for those considered on the $\PZ\PZ/\PW\PZ$ ratio in the $\PZ(\ell\ell)\PH$ channel, are correlated across all data sets.

\begin{figure*}[htb]
\centering
\includegraphics[width=0.45\textwidth]{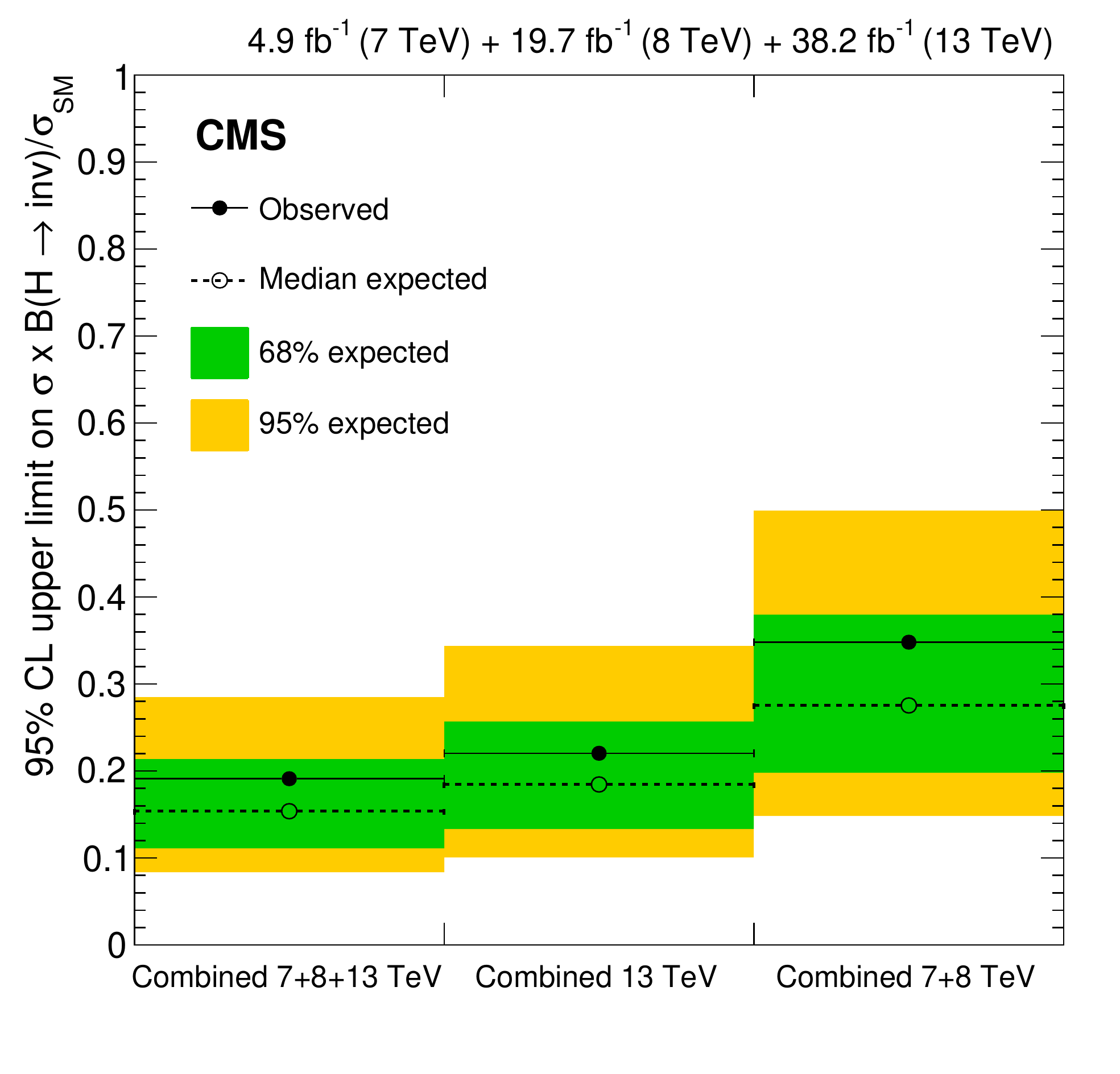}
\includegraphics[width=0.45\textwidth]{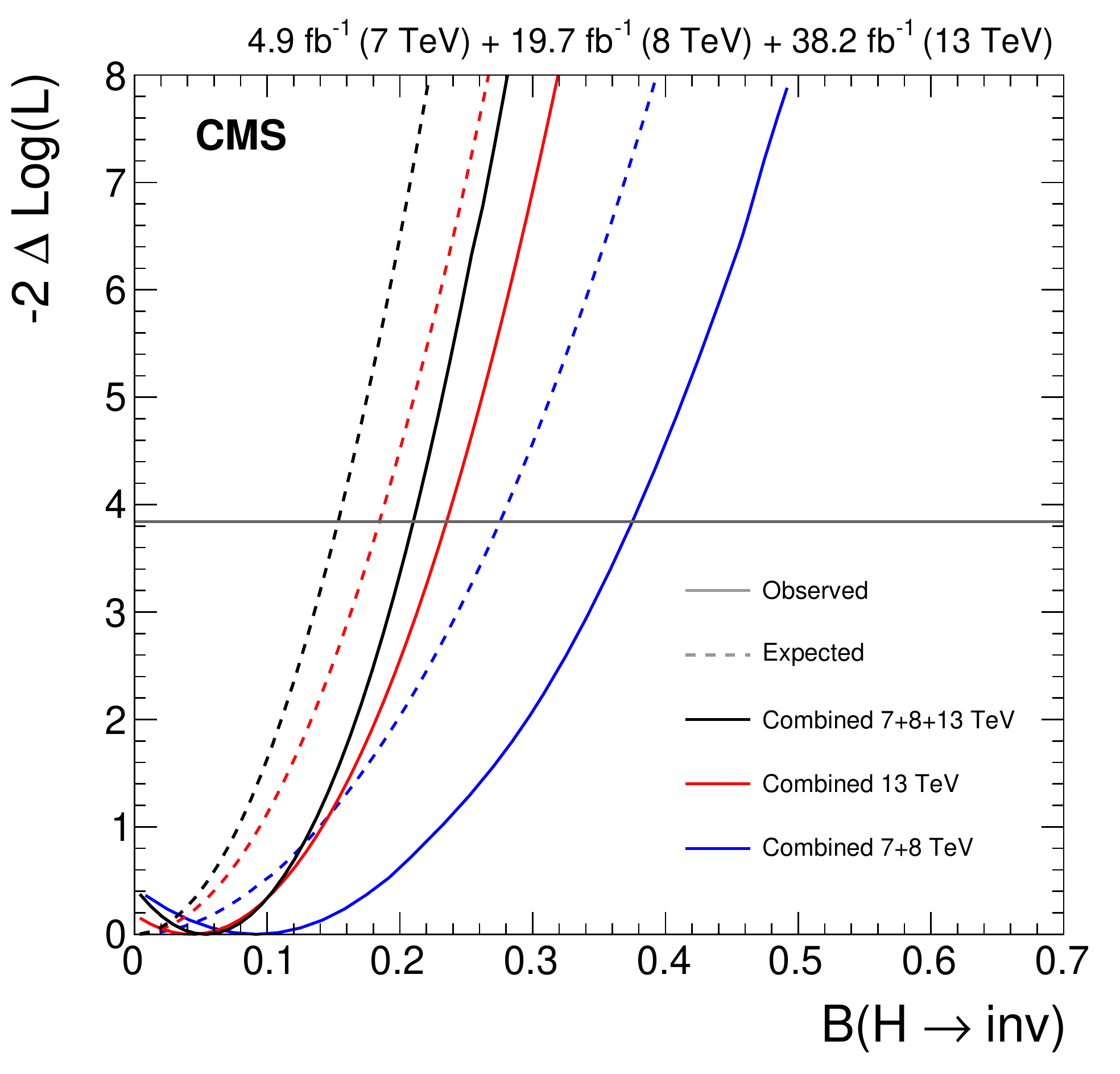}
\caption{On the left, observed and expected 95\% \CL upper limits on ${\sigmabr}$ for partial combinations based either on 7+8 or 13\TeV data as well as their combination, assuming SM production cross sections for the Higgs boson with mass of 125.09\GeV. On the right, the corresponding profile likelihood ratios as a function of \brhinv are presented. The solid curves represent the observations in data, while the dashed lines represent the expected result obtained from the background-only hypothesis.}
\label{fit:combination_direct_limit}
\end{figure*}

Observed and expected upper limits on ${\sigmabr}$ at 95\% \CL are presented in Fig.~\ref{fit:combination_direct_limit} (left). Limits are computed for the combination of all data sets, as well as for partial combinations based either on 7+8 or 13\TeV data. The relative contributions from different Higgs production mechanisms are constrained to their SM values within the theoretical uncertainties. The combination yields an observed (expected) upper limit of ${\brhinv < 0.19\,(0.15)}$ at 95\% \CL. The corresponding profile likelihood ratios as a function of \brhinv are shown in Fig.~\ref{fit:combination_direct_limit} (right). The measured value of the invisible branching fraction and an approximate 68\% \CL interval, obtained from the profile likelihood, are $\brhinv = 0.05 \pm 0.03\stat\pm 0.07\syst$. The systematic uncertainties with the highest impact in the \brhinv measurement are the theoretical uncertainties affecting the \Zvv/\Wlv and $\PZ\PZ/\PW\PW$ ratios in the VBF and $\PZ(\ell\ell)\PH$ channels, respectively, as well as the uncertainties in the lepton and photon reconstruction and identification efficiencies, jet energy scale, and veto efficiency of \tauh candidates.

\begin{figure*}[htb]
\centering
\includegraphics[width=0.45\textwidth]{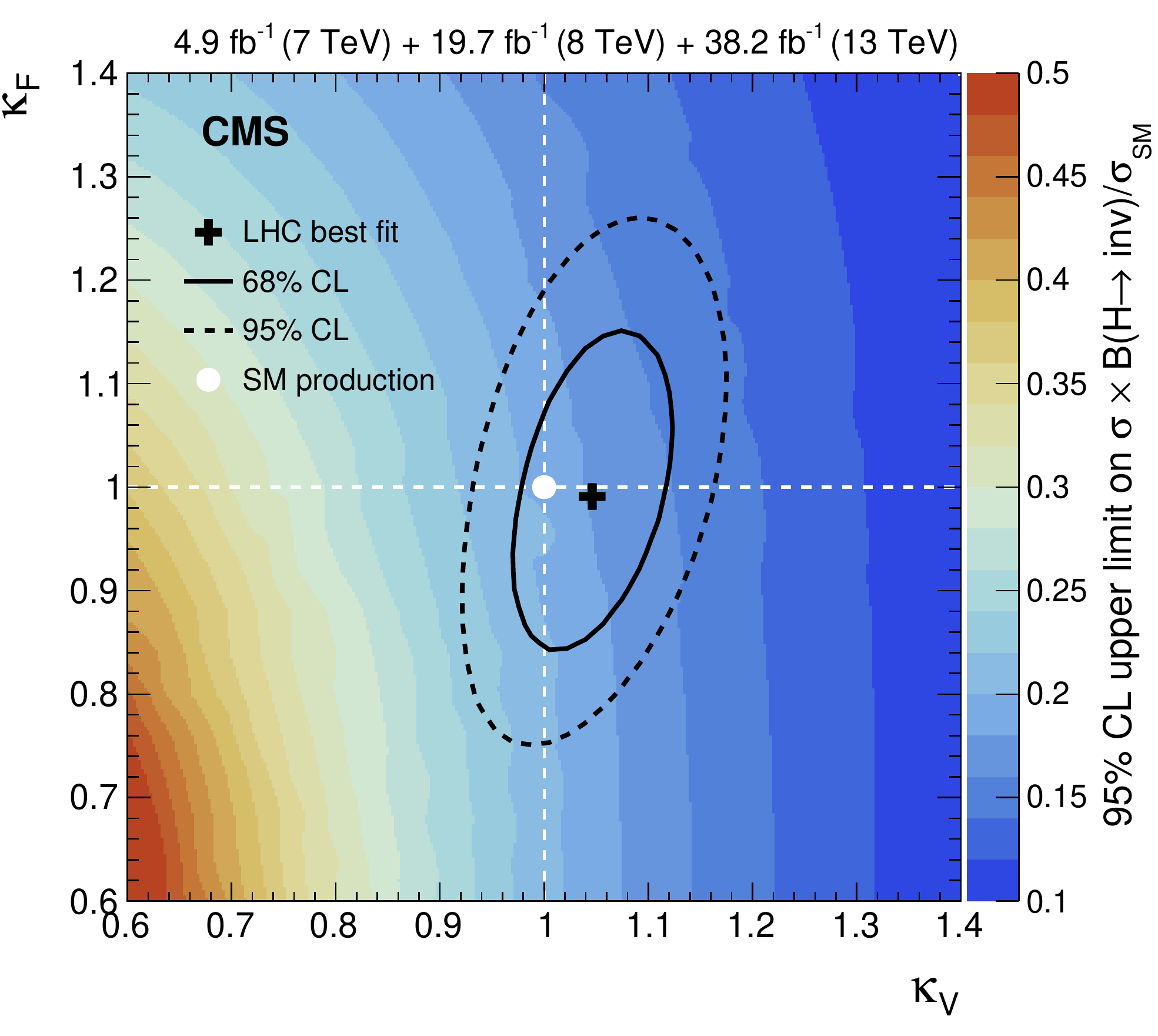}
\includegraphics[width=0.45\textwidth]{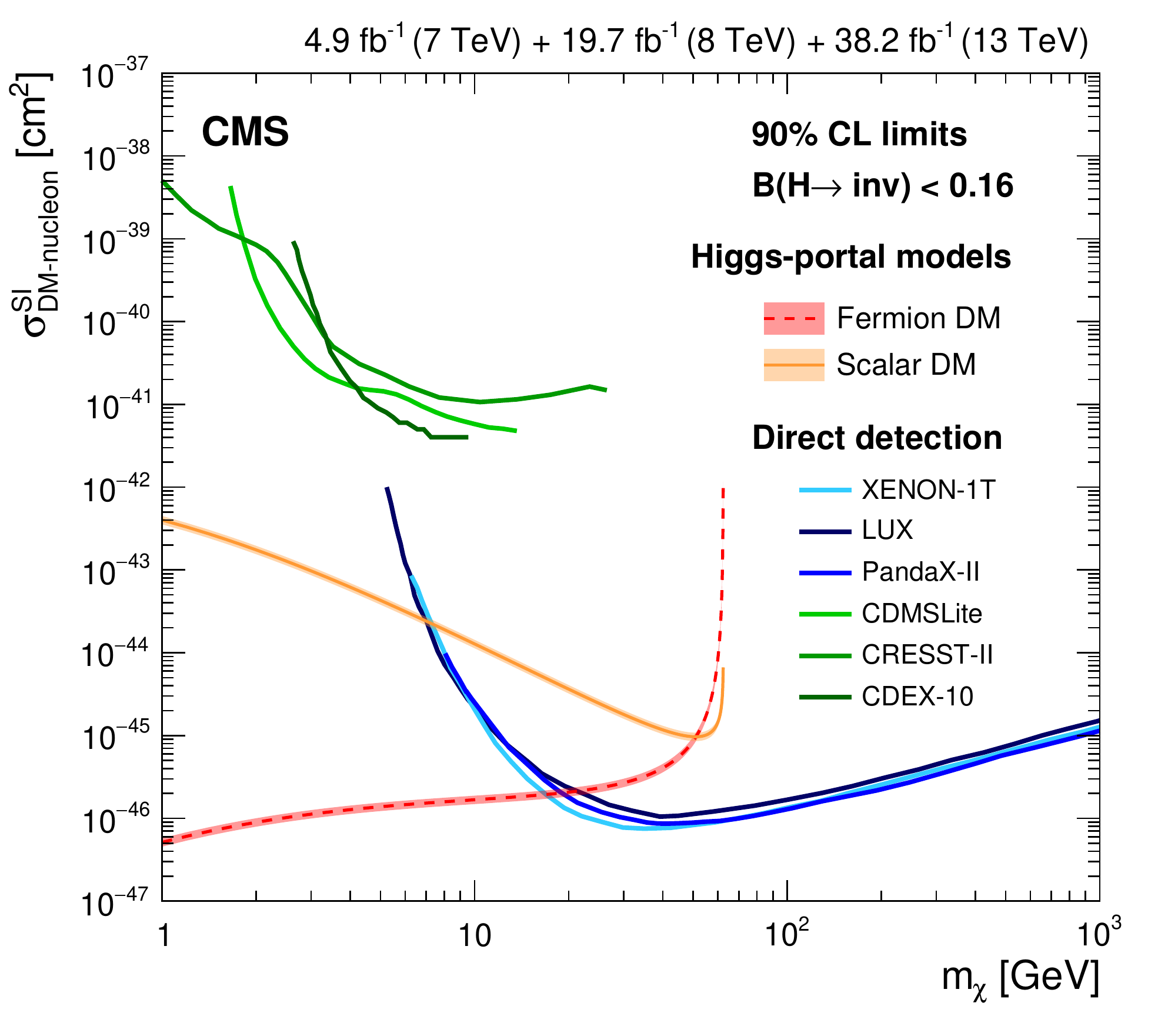}
\caption{On the left, observed 95\% \CL upper limits on \sigmabr for a Higgs boson with a mass of 125.09\GeV, whose production cross section varies as a function of the coupling modifiers $\kappa_{\mathrm{V}}$ and $\kappa_{\mathrm{F}}$. Their best estimate, along with the 68\% and 95\% \CL contours from Ref.~\cite{Khachatryan:2016vau}, are also reported. The SM prediction corresponds to ${\kappa_{\mathrm{V}} = \kappa_{\mathrm{F}} = 1}$. On the right, 90\% \CL upper limits on the spin-independent DM-nucleon scattering cross section in Higgs-portal models, assuming a scalar (solid orange) or fermion (dashed red) DM candidate. Limits are computed as a function of $m_{\chi}$ and are compared to those from the XENON1T~\cite{Aprile:2018dbl}, LUX~\cite{Akerib:2016vxi}, PandaX-II~\cite{Tan:2016zwf}, CDMSlite~\cite{Agnese:2015nto}, CRESST-II~\cite{Angloher:2015ewa}, and CDEX-10~\cite{Jiang:2018pic} experiments.}
\label{fig:bsmresults}
\end{figure*}

The relative sensitivity of each search considered in the combination depends on the assumed SM production rates. The cross sections for the $\Pg\Pg\PH$, VBF and $\PV\PH$ production modes are parametrized in terms of coupling strength modifiers $\kappa_{\mathrm{V}}$ and $\kappa_{\mathrm{F}}$, which directly scale the coupling of the Higgs boson to vector bosons and fermions, respectively~\cite{Heinemeyer:2013tqa}. The contribution from the ${\Pg\Pg \to \PZ\PH}$ production is scaled to account for the interference between the $\cPqt\PH$ and $\PZ\PH$ diagrams, as described in Ref.~\cite{deFlorian:2016spz}. In this context, SM production rates are obtained for ${\kappa_{\mathrm{V}} = \kappa_{\mathrm{F}} = 1}$. Figure~\ref{fig:bsmresults} (left) shows the observed 95\% \CL upper limits on ${\sigmabr}$ evaluated as a function of $\kappa_{\mathrm{V}}$ and $\kappa_{\mathrm{F}}$. The LHC best estimates for $\kappa_{\mathrm{V}}$ and $\kappa_{\mathrm{F}}$ from Ref.~\cite{Khachatryan:2016vau} are superimposed, along with the 68\% and 95\% \CL limit contours. Within the 95\% \CL region, the observed (expected) upper limit on \brhinv varies between 0.14\,(0.11) and 0.24\,(0.19).

The upper limit on \brhinv, obtained from the combination of $\sqrt{s}=7,$ 8, and 13\TeV searches, is interpreted in the context of Higgs-portal models of DM interactions, in which a stable DM particle couples to the SM Higgs boson. Direct-detection experiments are sensitive to the interaction between a DM particle and an atomic nucleus, which may be mediated by the exchange of a Higgs boson, producing nuclear recoil signatures that can be interpreted in terms of the DM-nucleon scattering cross section. The sensitivity of these experiments depends mainly on the DM particle mass ($m_{\chi}$). If $m_{\chi}$ is smaller than half of the Higgs boson mass, the Higgs boson invisible width ($\Gamma_{\text{inv}}$) can be translated, within an effective field theory approach, into a spin-independent DM-nucleon elastic scattering cross section, as outlined in Ref.~\cite{Djouadi:2011aa}. This translation is performed assuming that the DM candidate is either a scalar or a Majorana fermion, and both the central value and the uncertainty of the dimensionless nuclear form-factor $f_{N}$ are taken from the recommendations of Ref.~\cite{Hoferichter:2017olk}. The conversion from \brhinv to $\Gamma_{\text{inv}}$ uses the relation ${\brhinv = \Gamma_{\text{inv}} / (\Gamma_{\mathrm{SM}}+\Gamma_{\text{inv}})}$, where $\Gamma_{\mathrm{SM}}$ is set to 4.07\MeV~\cite{Heinemeyer:2013tqa}. Since renormalizable models predicting a vectorial DM candidate require an extended dark Higgs sector, which may lead to modifications of kinematic distributions assumed for the invisible Higgs boson signal, such interpretation is not provided in the context of this Letter. Figure~\ref{fig:bsmresults} (right) shows the 90\% \CL upper limits on the spin-independent DM-nucleon scattering cross section as a function of $m_{\chi}$, for both the scalar and the fermion DM scenarios. These limits are computed at 90\% \CL so that they can be compared with those from direct detection experiments such as XENON1T~\cite{Aprile:2018dbl}, LUX~\cite{Akerib:2016vxi}, PandaX-II~\cite{Tan:2016zwf}, CDMSlite~\cite{Agnese:2015nto}, CRESST-II~\cite{Angloher:2015ewa}, and CDEX-10~\cite{Jiang:2018pic} which provide the strongest constraints in the $m_{\chi}$ range probed by this search. In the context of Higgs-portal models, the result presented in this Letter provides the most stringent limits for $m_{\chi}$ smaller than 18\,(7)\GeV, assuming a fermion (scalar) DM candidate.

\section{Summary}\label{sec:summary}

A search for invisible decays of a Higgs boson is presented using proton-proton ($\Pp\Pp$) collision data at a center-of-mass energy $\sqrt{s} = 13\TeV$, collected by the CMS experiment in 2016 and corresponding to an integrated luminosity of 35.9\fbinv. The search targets events in which a Higgs boson is produced through vector boson fusion (VBF). The data are found to be consistent with the predicted standard model (SM) backgrounds. An observed (expected) upper limit of 0.33\,(0.25) is set, at 95\% confidence level (\CL), on the branching fraction of the Higgs boson decay to invisible particles, \brhinv, by means of a binned likelihood fit to the dijet mass distribution. In addition, upper limits are set on the product of the cross section and branching fraction of an SM-like Higgs boson, with mass ranging between 110 and 1000\GeV.

A combination of CMS searches for the Higgs boson decaying to invisible particles, using $\Pp\Pp$ collision data collected at $\sqrt{s} = 7,$ 8, and 13\TeV (2015 and 2016), is also presented. The combination includes searches targeting Higgs boson production via VBF, in association with a vector boson (with hadronic decays of the $\PW$ boson and hadronic or leptonic decays of the $\PZ$ boson) and via gluon fusion with initial state radiation. The VBF search is the most sensitive channel involved in the combination. No significant deviations from the SM predictions are observed in any of these searches. The combination yields an observed (expected) upper limit on \brhinv of 0.19\,(0.15) at 95\% \CL, assuming SM production rates for the Higgs boson and a Higgs boson mass of 125.09\GeV. The observed 90\% \CL upper limit of $\brhinv < 0.16$ is interpreted in terms of Higgs-portal models of dark matter (DM) interactions. Constraints are placed on the spin-independent DM-nucleon interaction cross section. When compared to the upper bounds from direct detection experiments, this limit provides the strongest constraints on fermion (scalar) DM particles with masses smaller than about 18\,(7)\GeV.

\begin{acknowledgments}

We congratulate our colleagues in the CERN accelerator departments for the excellent performance of the LHC and thank the technical and administrative staffs at CERN and at other CMS institutes for their contributions to the success of the CMS effort. In addition, we gratefully acknowledge the computing centres and personnel of the Worldwide LHC Computing Grid for delivering so effectively the computing infrastructure essential to our analyses. Finally, we acknowledge the enduring support for the construction and operation of the LHC and the CMS detector provided by the following funding agencies: BMWFW and FWF (Austria); FNRS and FWO (Belgium); CNPq, CAPES, FAPERJ, and FAPESP (Brazil); MES (Bulgaria); CERN; CAS, MoST, and NSFC (China); COLCIENCIAS (Colombia); MSES and CSF (Croatia); RPF (Cyprus); SENESCYT (Ecuador); MoER, ERC IUT, and ERDF (Estonia); Academy of Finland, MEC, and HIP (Finland); CEA and CNRS/IN2P3 (France); BMBF, DFG, and HGF (Germany); GSRT (Greece); NKFIA (Hungary); DAE and DST (India); IPM (Iran); SFI (Ireland); INFN (Italy); MSIP and NRF (Republic of Korea); LAS (Lithuania); MOE and UM (Malaysia); BUAP, CINVESTAV, CONACYT, LNS, SEP, and UASLP-FAI (Mexico); MBIE (New Zealand); PAEC (Pakistan); MSHE and NSC (Poland); FCT (Portugal); JINR (Dubna); MON, RosAtom, RAS and RFBR (Russia); MESTD (Serbia); SEIDI, CPAN, PCTI and FEDER (Spain); Swiss Funding Agencies (Switzerland); MST (Taipei); ThEPCenter, IPST, STAR, and NSTDA (Thailand); TUBITAK and TAEK (Turkey); NASU and SFFR (Ukraine); STFC (United Kingdom); DOE and NSF (USA).

\hyphenation{Rachada-pisek} Individuals have received support from the Marie-Curie programme and the European Research Council and Horizon 2020 Grant, contract No. 675440 (European Union); the Leventis Foundation; the A. P. Sloan Foundation; the Alexander von Humboldt Foundation; the Belgian Federal Science Policy Office; the Fonds pour la Formation \`a la Recherche dans l'Industrie et dans l'Agriculture (FRIA-Belgium); the Agentschap voor Innovatie door Wetenschap en Technologie (IWT-Belgium); the F.R.S.-FNRS and FWO (Belgium) under the ``Excellence of Science - EOS'' - be.h project n. 30820817; the Ministry of Education, Youth and Sports (MEYS) of the Czech Republic; the Lend\"ulet (``Momentum'') Programme and the J\'anos Bolyai Research Scholarship of the Hungarian Academy of Sciences, the New National Excellence Program \'UNKP, the NKFIA research grants 123842, 123959, 124845, 124850 and 125105 (Hungary); the Council of Science and Industrial Research, India; the HOMING PLUS programme of the Foundation for Polish Science, cofinanced from European Union, Regional Development Fund, the Mobility Plus programme of the Ministry of Science and Higher Education, the National Science Center (Poland), contracts Harmonia 2014/14/M/ST2/00428, Opus 2014/13/B/ST2/02543, 2014/15/B/ST2/03998, and 2015/19/B/ST2/02861, Sonata-bis 2012/07/E/ST2/01406; the National Priorities Research Program by Qatar National Research Fund; the Programa Estatal de Fomento de la Investigaci{\'o}n Cient{\'i}fica y T{\'e}cnica de Excelencia Mar\'{\i}a de Maeztu, grant MDM-2015-0509 and the Programa Severo Ochoa del Principado de Asturias; the Thalis and Aristeia programmes cofinanced by EU-ESF and the Greek NSRF; the Rachadapisek Sompot Fund for Postdoctoral Fellowship, Chulalongkorn University and the Chulalongkorn Academic into Its 2nd Century Project Advancement Project (Thailand); the Welch Foundation, contract C-1845; and the Weston Havens Foundation (USA).

\end{acknowledgments}

\bibliography{auto_generated}

\cleardoublepage \appendix\section{The CMS Collaboration \label{app:collab}}\begin{sloppypar}\hyphenpenalty=5000\widowpenalty=500\clubpenalty=5000\vskip\cmsinstskip
\textbf{Yerevan Physics Institute, Yerevan, Armenia}\\*[0pt]
A.M.~Sirunyan$^{\textrm{\dag}}$, A.~Tumasyan
\vskip\cmsinstskip
\textbf{Institut f\"{u}r Hochenergiephysik, Wien, Austria}\\*[0pt]
W.~Adam, F.~Ambrogi, T.~Bergauer, J.~Brandstetter, M.~Dragicevic, J.~Er\"{o}, A.~Escalante~Del~Valle, M.~Flechl, R.~Fr\"{u}hwirth\cmsAuthorMark{1}, M.~Jeitler\cmsAuthorMark{1}, N.~Krammer, I.~Kr\"{a}tschmer, D.~Liko, T.~Madlener, I.~Mikulec, N.~Rad, J.~Schieck\cmsAuthorMark{1}, R.~Sch\"{o}fbeck, M.~Spanring, D.~Spitzbart, W.~Waltenberger, J.~Wittmann, C.-E.~Wulz\cmsAuthorMark{1}, M.~Zarucki
% ...
\vskip\cmsinstskip
\dag: Deceased\\
1:  Also at Vienna University of Technology, Vienna, Austria\\
% ...
\end{sloppypar}
\end{document}